\documentclass[12pt,preprint,aps,prb,floatfix]{revtex4}
\usepackage{amsmath}
\usepackage{graphicx}

\begin{document}

\title{Equation of State for Macromolecules of Variable Flexibility in Good Solvents:
 A Comparison of Techniques for Monte Carlo Simulations of Lattice Models}

\author{V. A. Ivanov, E. A. An, L. A. Spirin,
M. R. Stukan,\footnote{Current address: Schlumberger Cambridge Research, High Cross, Madingley
Road, Cambridge CB3 OEL, UK}}
\affiliation{Faculty of Physics, Moscow State University, Moscow 119992, Russia}

\author{M. M\"uller, }
\affiliation{Institut f\"ur Theoretische Physik,
Georg-August-Universit\"at, Friedrich-Hund-Platz 1, 37077
G\"ottingen, Germany}

\author{W. Paul, and K. Binder}
\affiliation{Institut f\"ur Physik,
Johannes-Gutenberg-Universit\"at, Staudinger Weg 7, 55099 Mainz,
Germany}

\date{\today}

\begin{abstract}
The osmotic equation of state for the athermal bond fluctuation
model on the simple cubic lattice is
obtained from extensive Monte Carlo simulations. For short macromolecules (chain
length N=20) we study the influence of various choices for the chain
stiffness on the equation of state. Three techniques
are applied and compared in order to critically assess their
efficiency and accuracy: the ``repulsive wall'' method, the
thermodynamic integration method (which rests on the feasibility
of simulations in the grand canonical ensemble), and the recently
advocated sedimentation equilibrium method, which records the
density profile in an external (e.g. gravitation-like) field and
infers, via a local density approximation, the equation of state
from the hydrostatic equilibrium condition. We confirm the
conclusion that the latter technique is far more efficient than
the repulsive wall method, but we find that the thermodynamic
integration method is similarly efficient as the sedimentation
equilibrium method. For very stiff chains the onset of nematic order
enforces the formation of an isotropic--nematic interface in the
sedimentation equilibrium method leading to strong rounding effects
and deviations from the true equation of state in the transition
regime.

\end{abstract}

\maketitle

\section{Introduction}
\label{introduction}

Understanding the equation of state of macromolecules in solution
has been a longstanding problem of polymer science, which is
both important as a fundamental problem in the statistical
mechanics of soft
matter~\cite{Flory,deGennes,Cloizeaux,Grosberg,Binder,Schafer,Rubinstein}
and relevant for various applications of polymers. The
interplay of excluded volume
effects~\cite{deGennes,Cloizeaux,Grosberg,Schafer}, solvent
quality and variable chain stiffness~\cite{Grosberg,KhoSem} already
is very difficult to describe for macromolecules in very dilute
solution~\cite{Seeeg}, withstanding an analytic solution and making
the application of computer simulation
methods~\cite{Binder,Kotelyanskii} necessary. The experimental
situation on measurements of the equation of state of polymer
solutions is addressed in chapter $5$ of Ref.~\onlinecite{Cloizeaux}
and chapter $3$ of Ref.~\onlinecite{Strobl}. Considering
semidilute and concentrated solutions, the enthalpic and/or
entropic interactions among the polymer chains create nontrivial
correlations in the structure of the solutions involving many
chains, and phase transitions such as phase separation under bad
solvent conditions into a dilute polymer solution and a concentrated
one may occur~\cite{Flory,deGennes,Widom,CBGC,Binder2}.
Alternatively, under good solvent conditions one may observe for
semiflexible or for stiff chains a transition from an isotropic to
a nematic solution when the polymer concentration
increases~\cite{Flory2,Baumgartner,Weber,Strey,Vega}. Of
particular interest is the situation when nematic ordering and the
tendency to phase separation under bad solvent conditions
compete~\cite{Khokhlov}. Preliminary computer simulation studies
of a corresponding model~\cite{LCpaper1,LCpaper2} gave only rather
rough information on the phase behavior, and it was concluded that the accuracy with
which the osmotic equation of state could be determined needs
to be improved.

Thus, there are many reasons why establishing methods that allow
the reliable estimation of the equation of state for various
coarse-grained lattice models for polymers is of interest. While
in an off-lattice model the estimation of the pressure tensor from
the virial theorem in principle is straightforward~\cite{Allen},
it often is not possible to study very large systems with the
desirable accuracy~\cite{Binder}, and hence simulations of lattice
models still have their place~\cite{Binder,Kotelyanskii}. However, for lattice
models different techniques to calculate the osmotic pressure of
polymers in solution must be sought in order to derive their
equation of state. The standard method, particularly valuable for
dilute solutions and/or not too large chain lengths, calculates
the chemical potential from the insertion probability of a test
chain~\cite{Bell,latt1eos} and the osmotic pressure then follows
from standard thermodynamic
integration~\cite{Binder,Allen,Frenkel,Landau}. As is well known,
insertion of a polymer chain into a volume containing already
other chains is very difficult due to very small acceptance
probabilities, and hence requires advanced Monte Carlo
methods~\cite{Frenkel,Siepmann,MuellerPaul} to be practically feasible.

As an alternative method Dickman~\cite{rep_wall1, rep_wall2} proposed the
repulsive wall thermodynamic integration (RWTI) method, which
remains applicable also for very dense systems since it works in
the canonic (${\cal N}VT$) ensemble, where the number of chains $\cal N$ in the
box remains fixed. However, this method requires substantial
simulation effort, since for each state point in the bulk several
simulation runs with increasing wall--monomer repulsion and
subsequent thermodynamic integration need to be performed.
Moreover the RWTI method may suffer from important finite size
effects~\cite{PressPaper}.

More recently, it was proposed~\cite{Hansen} to extract the
osmotic equation of state from Monte Carlo simulations where the
equilibrium monomer and center-of-mass concentration profiles of
lattice polymers in a gravitation-like potential are computed.
From these concentration profiles, the equation of state can be
inferred if a local density approximation like in hydrostatic
equilibrium is invoked. This method has been broadly applied to
study the sedimentation equilibrium of colloidal dispersions
containing spherical~\cite{BibenHansen},
rod-like~\cite{Dijkstra04} or disk-like~\cite{DijkstraHansen}
particles. More recently, successful applications to
colloid-polymer mixtures~\cite{HansenJPCM} and solutions of block
copolymers~\cite{HansenMolPhy} or binary polymer
solutions~\cite{HansenJPCB} have been made as well.

Despite these successes, this ``sedimentation equilibrium'' (SE)
method also may have drawbacks, when other large length scales appear in
the system, that compete with the characteristic ``sedimentation
length''. This length that scales inversely with the ``gravitational
constant'' characterizing the gravitation-like potential and hence
can be made arbitrarily large~\cite{Hansen}, but the linear
dimension of the simulated system in the $z$-direction in which
this gravitation-like force acts must then be huge, also.
A well-known case, where even the real gravitation potential on
earth substantially disturbs the equation of state is a fluid very
close to the gas-liquid critical point~\cite{HohenBarm}, since
there the correlation length of density fluctuations diverges, and
critical fluctuations undisturbed by gravity can occur in the
$x,y$-directions perpendicular to the gravitational force
only~\cite{HohenBarm}. Other cases, apart from critical points of
second-order transitions, where large length scales arise that are
potentially disturbed by the gravitation-like potential are
associated with the formation of thick wetting layers at walls,
for instance.

While Addison at al.~\cite{Hansen} did test the SE method against
the RWTI method for a few cases varying the solvent quality from
good solvents to theta solvents, finding good agreement, they
considered only fully flexible chains. Thus, we take up this
problem but rather consider semiflexible chains as well: the
possible occurrence of nematically ordered wetting layers at the
wall where the density is largest~\cite{Dijkstra03} or even the
occurrence of the isotropic to nematic transition in the bulk
solution of the semiflexible or stiff polymers may complicate
matters. The aim of the present work is to carefully test the SE
method against both the RWTI method and the standard thermodynamic
integration method in the grand canonical $\mu VT$ ensemble
(TI$\mu VT$ method, $\mu$ denoting the chemical potential), and to
assess by detailed comparisons both the efficiency and the
accuracy of these methods. As has been explained above, there is need
for accurate simulation methods for the computation of the
equation of state of polymer solutions in the context of various
interesting problems.

The plan for the remainder of this paper is as follows. In
Sec.~\ref{methods} we briefly review the theoretical basis for the
RWTI, TI$\mu VT$ and SE methods, while Sec.~\ref{model} describes
the simulated model and mentions the types of Monte Carlo moves
used. Sec.~\ref{flexibleEOS} presents comparisons of the three
methods for fully flexible chains over a wide range of densities,
while Sec.~\ref{stiffEOS} presents our results for variable chain
stiffness. Sec.~\ref{conclusions} contains our conclusions and
gives an outlook to future work.

\section{Methods to calculate the osmotic pressure for lattice models of polymer solutions}
\label{methods}

\subsection{Thermodynamic integration in the grand canonical ensemble (TI$\mu VT$ method)}

We consider a system of chains of length $N$ in a simulation box of volume
$V$ with periodic boundary conditions (typically the box has a cubic shape,
$V=L^3$, $L$ being  the linear dimension) at given temperature $T$ and chemical potential
$\mu$ of the chains. The density of polymer chains in the system, $\rho={\cal N}/V$, with ${\cal N}$ the
average number of chains contained in the simulation box then is an output
of the simulation.

Utilizing the standard relations in the canonical ensemble, $\mu=(\partial(F/V)/\partial\rho)_{T,V}$
and $p=-(\partial F/\partial V)_{T,{\cal N}}$, where $F$ is the free energy of the system and $p$ the
pressure, one easily derives that
\begin{equation}
p=\rho\mu-\int\limits^{\rho}_0\mu(\rho')d\rho'+const,
\label{press}
\end{equation}
where the integration constant in Eq.~(\ref{press}) can be fixed by reference to the low density
limit, where the system behaves like an ideal gas of chains,
\begin{equation}
pV={\cal N}k_BT, \quad \pi\equiv p/k_BT=\rho.
\label{idealgas}
\end{equation}

Denoting then the chemical potential of the ideal gas of chains as $\mu_{id}$, and defining
$\mu^{ex}=\mu-\mu_{id}$ the excess chemical potential per chain, Eqs.~(\ref{press}),(\ref{idealgas}) imply~\cite{Bell}
\begin{equation}
\pi=\rho(1+\mu^{ex})-\int\limits_0^{\rho}\mu^{ex}(\rho')d\rho'
\label{osmpr}
\end{equation}
In practice the integral in Eq.~(\ref{osmpr}) is discretized, so that the reduced osmotic pressure $\pi_i$
at chain density $\rho_i$ is obtained from the recursion relation
\begin{equation}
\pi_i\approx\pi_{i-1}+(1+\mu_i^{ex})\rho_i-(1+\mu_{i-1}^{ex})\rho_{i-1}-(\mu^{ex}_i+\mu^{ex}_{i-1})(\rho_i-\rho_{i-1})/2.
\label{pii}
\end{equation}
The disadvantages of the method are clearly obvious from Eq.~(\ref{pii}): a large number
of state points $\{\mu_i,T,V\}$ needs to be studied with small differences between $\mu_i$ and $\mu_{i-1}$
and hence $\rho_i$ and $\rho_{i-1}$, so that the discretization error going
from Eq.~(\ref{osmpr}) to Eq.~(\ref{pii}) is negligible; and an efficient grand canonical simulation method is
needed, so $\rho_i$ is sampled with sufficient accuracy. For not too long chains and not too
high densities, however, the method is indeed practically useful, and since periodic
boundary conditions are used, the system for large $V$ always is homogeneous, and
the analysis is not hampered by any interfacial effects, which are present inevitably
in the other techniques described below due to their explicit use of walls.

\subsection{The repulsive wall thermodynamic integration (RWTI) method}

This method can be implemented both in the grand canonical $\mu VT$ ensemble and
in the canonical ${\cal N}VT$ ensemble. While the latter choice has the advantage that
no chain insertions are necessary and hence the method also works for long chains
and at high densities, it suffers from rather large finite size effects~\cite{PressPaper}. We
have used here both the canonical and the grand canonical version of the method.

One considers now a box of linear dimensions $L\times L\times H$, with periodic
boundary conditions in $x$ and $y$ directions only, while a hard wall is placed
both at $z=0$ and $z=H$. Moreover, one introduces a repulsive potential
of strength $\varepsilon_{wall}$ which acts in the first layer adjacent to $z=0$ and in the last
layer available to the monomers, $z=H$. As discussed in~\cite{rep_wall1,rep_wall2,PressPaper}, the osmotic
pressure is then obtained from the fraction of sites, $\phi_z(\lambda)$, occupied by monomers
in the layers adjacent to both walls, $z=0$ and $z=H$,

\begin{equation}
\pi=\int\limits_0^1\frac{d\lambda}{\lambda}
\left(\frac{\phi_{z=0}(\lambda)+\phi_{z=H}(\lambda)}{2}\right), \quad
\lambda\equiv\exp(-\varepsilon_{wall}/k_BT).
\label{OsmRWM}
\end{equation}

Again the integration in Eq.~(\ref{OsmRWM}) is discretized, and 20 different values of $\lambda$
turned out to be sufficient to obtain reliable results.

\subsection{The sedimentation equilibrium (SE) method}

This method utilizes the canonical ensemble ${\cal N}VT$, and one also considers
a box of linear dimensions $L\times L \times H$ with periodic boundary conditions in $x$ and
$y$ directions only, while again hard walls are used in $z$-direction at $z=0$ and
at $z=H$. An external potential is applied not at the walls but everywhere in the
system~\cite{Hansen}

\begin{equation}
U_{\rm external}(z)=-mgz=-\frac{\displaystyle k_BT}{\displaystyle a}\lambda_gz.
\label{ExtPot}
\end{equation}

Here $m$ is the mass of a monomer, $g$ is the acceleration due to the gravity-like
potential, $a$ is the lattice spacing,
and $\lambda_g$ is a dimensionless constant that
characterizes the strength of this gravitational potential, $\lambda_g=amg/k_BT$.
It is also useful to introduce characteristic gravitational lengths $\xi_m,\xi_{cm}$
\begin{equation}
\xi_m = a/\lambda_g, \quad \xi_{cm} = \xi_m/N.
\label{grlength}
\end{equation}
As shown by Addison et al.~\cite{Hansen}, for large $z$ the density profile of an
ideal gas of monomers at the lattice would follow the standard
barometric formula, $\rho_m(z)\propto \exp(-mgz/k_BT)=\exp(-z/\xi_m)$, while
for an ideal gas of polymer chains the density profile of monomer units
$\rho_m(z)=N\rho(z)\propto\exp(-z/\xi_{cm})$.

While the variation of the density profile for large $z$, where the
system is very dilute and the ideal gas behavior holds, hence is trivially
known, and this knowledge is an important consistency check of the method,
for smaller $z$ the density profile is nontrivial. But from this profile
the osmotic equation of state can be estimated, when one invokes the local
density approximation, such that the equation of hydrostatic equilibrium
holds~\cite{Hansen,BibenHansen}
\begin{equation}
\frac{\displaystyle dp(z)}{\displaystyle dz}= - Nmg\rho(z),
\label{Osmloc}
\end{equation}
$p(z)$ being the local osmotic pressure at altitude $z$. Integration of Eq.(\ref{Osmloc})
yields
\begin{equation}
\pi(z)=p(z)/k_BT=\xi^{-1}_{cm}\int\limits^{\infty}_{z}\rho(z')dz'.
\label{OsmSE}
\end{equation}

Thus one can record both $\rho(z)$ (and the associated monomer profile
$\rho_m(z)$) and $\pi(z)$ simultaneously, and eliminating $z$ from these relations
one obtains the desired equation of state $\pi (\rho)$ (or $\pi (\rho_m)$, respectively).

Noting that in the local density approximation we expect that the density
distribution of the monomer units satisfies $\rho_m(z)=N\rho(z)$,
we find from Eqs.~(\ref{grlength}), (\ref{OsmSE})
that $\pi(z)$ can also be written as
\begin{equation}
\pi(z)=\xi^{-1}_m\int\limits^{\infty}_{z}\rho_m(z')dz'.
\label{OsmSE2}
\end{equation}

However, the validity of the local density approximation needs to be considered
carefully. E.g., near the hard wall at $z=0$, $\rho_m(z)$ may exhibit strong
oscillations (``layering''), and also $\rho(z)$ exhibits a nontrivial structure.
Thus it is clear that the local density approximation breaks down near the
hard wall at $z=0$, and in fact one should use Eqs.~(\ref{OsmSE}),(\ref{OsmSE2}) only for $z\ge R_g$,
the gyration radius of the chains. Similarly, rapid density variations invalidating
the local density approximation are also expected near an interface between coexisting
phases (as it may occur for bad solvent conditions, for instance). In addition,
one must require that on the scale of $\Delta z = R_g$ the change of the potential,
Eq.~(\ref{ExtPot}), is negligibly small. This implies
\begin{equation}
\left| \Delta U_{\rm external}(\Delta z)/k_BT \right|=\lambda_gR_g/a=R_g/\xi_m\ll 1.
\label{DExtPot}
\end{equation}
For flexible chains, $R_g$ scales like $\sqrt{N}$ in concentrated solutions, while for stiff
chains we have $R_g\propto N$. This implies that for stiff chains considerably larger
values of $\xi_m$ (and hence smaller values of $\lambda_g$) need to be chosen than for flexible
ones. Of course, since one must accommodate in the simulation box the full
density profile from a value appropriate for concentrated solutions near $z=0$
down to the dilute regime near $z=H$, a linear dimension $H\gg\xi_m$ is mandatory;
otherwise, the distortion of the density profile due to this upper boundary at $z=H$
leads to systematic errors as well. Thus, the SE method requires a careful choice
of simulation parameters to avoid such errors and ensure the desired very good
accuracy, and the large size $H$ in $z$-direction to some extent will reduce the
advantage, that the whole equation of state can be estimated from a single run.

Finally we note that the approach can be generalized to other forms of
the external potential $U_{external}(z)$, different from a gravitation-like potential.
E.g., if one uses
\begin{equation}
U'_{\rm external}(z)=-\frac{\displaystyle k_BT}{\displaystyle a}\lambda'_gz^\kappa,
\label{ExtPotzk}
\end{equation}
where $\kappa$ is an exponent different from one, the equation for the force balance at
height $z$ \{Eq.~\ref{Osmloc}\} gets modified as
\begin{equation}
\frac{\displaystyle dp(z)}{\displaystyle dz}= - (\lambda'_g/a)\kappa\rho(z)z^{\kappa-1},
\label{Osmlock}
\end{equation}
and hence
\begin{equation}
\pi(z)=(\xi'_m)^{-1}\int\limits^{\infty}_{z}\kappa {\tilde z}^{\kappa-1}\rho_m({\tilde z})d{\tilde z}.
\label{OsmSEk}
\end{equation}

Motivation for using different potentials can be experiments made in a centrifuge.
The centrifugal force depends linearly on the distance from the center of rotation,
so that $\kappa=2$ for this case.
Although it is not clear at this point which choice for the exponent $\kappa$ is optimal,
testing that the equation of state $\pi(\rho_m)$ thus obtained does not depend on $\kappa$
is a nice consistency check.

\section{Model and Simulation Technique}
\label{model}

We use the bond fluctuation model~\cite{BFM} on the simple cubic lattice, taking
henceforth $a\equiv 1$ as our unit of length. Each effective monomeric unit is represented
by an elementary cube of the lattice, blocking all 8 sites at the corners of this cube
from further occupation, realizing thus the excluded volume interaction between
the monomers. The bond vectors can be taken from the set $\left\{(\pm 2,0,0),(\pm 2,\pm 1,0),
(\pm 2,\pm 1,\pm 1), (\pm 2,\pm 2,\pm 1), (\pm 3,0,0), (\pm 3,\pm 1,0)\right\}$, including also all permutations
between these coordinates, -- altogether 108 different bond vectors occur, which leads
to 87 different angles between successive bonds. To model the chain stiffness, an
intramolecular potential depending on the angle $\vartheta$ between two successive bond
vectors along the chain is introduced (bending energy), and also an energy term
depending on the bond length $b$ can be included~\cite{Weber},
\begin{equation}
U_{\rm bending} = U_{\vartheta} + U_{b} =
-f \cos \vartheta (1 + c \cos \vartheta) + \varepsilon_0 \left(b-b_0\right)^2.
\label{Estiff}
\end{equation}
Both the stiffness parameter $f$ and $\varepsilon_0$ are measured in units of $k_BT$. Note that
for the model with $f=2.68$, $\varepsilon_0=4$, $b_0=0.86$, $c=0.03$ the isotropic to nematic transition
has been studied by extensive Monte Carlo simulations previously~\cite{Weber}, and in
order to be able to compare the present results to previous work we have included
this bond length energy term here again.

Note that variable solvent quality can be modeled by including also
a square well potential between any pair of monomers,
\begin{equation}
U_{\rm SW} (r)=
\begin{cases}
- \varepsilon, & r\leq\sqrt{6}\\
 0, & r>\sqrt{6},
\end{cases}
\label{Etherm}
\end{equation}
as done in Refs.~\onlinecite{CBGC,LCpaper1}.
However, the present work will treat the case $\varepsilon=0$ only, our
analysis of the phase behavior when both $f$ and $\varepsilon$ are nonzero is deferred to a
later work.

For the runs in the ${\cal N}VT$ ensemble (for the RWTI and SE methods)
we define two Monte Carlo steps (MCS) to involve one attempt to perform a local
``random hopping'' move per each monomer unit in the system~\cite{BFM} and one attempt
of a slithering-snake move per chain. These are the standard moves for simulations
using the bond fluctuation model~\cite{Binder,Binder3} in the canonical ensemble. The chain length
used was $N=20$ throughout, while typical box sizes were $90\times 90\times 90$ and
$60\times 60\times 180$ for the RWTI method and $80\times 80\times 250$ for the SE method, respectively.
For the SE method, a typical system contained ${\cal N}=1000$ chains,
and about $10^7$ MCS were needed to get a reasonably well equilibrated density
profile. For the RWTI method, the number of chains in the box was varied
from very small number up to about ${\cal N}=3000$. Again typically $10^7$ MCS per run
were used, taking $10^4$ ``measurements'' in each run, with $10^3$  MCS between two
successive measurements (it was checked that these $10^4$ configurations then were
uncorrelated).

For the runs in the $\mu VT$ ensemble, one Monte Carlo step means one
configurational bias move~\cite{CBGC,LCpaper1,Frenkel,Siepmann} plus additionally either one attempt to
perform a local random hopping move of every effective monomer in the system or
one attempt of a slithering-snake move per chain. For the TI$\mu VT$ method, about
50--60 different values of the chemical potential were used, and the box size was
$90\times 90\times 90$. For each parameter combination $3\cdot 10^5$ MCS were taken. The maximum
value of the number of chains reached was about ${\cal N}=2700$
(this corresponds to the polymer volume fraction
of about $\phi=0.6$ in this simulation box). So the total number of MCS needed to
record the equation of state is of the order of $2\cdot 10^7$ MCS (but note that 1 MCS
needs an amount of CPU time which depends on the number of chains  ${\cal N}$ in the system).

Since it is well known~\cite{Binder,Landau} that in the ${\cal N}VT$ ensemble the equilibration
of long wavelength density fluctuations is very slow, suffering from ``hydrodynamic
slowing down''~\cite{Landau}, it is useful to take special precautions that the density
profiles in the RWTI are well equilibrated, in order to avoid uncontrolled errors.
Therefore for each value of the repulsive wall parameter $\lambda$ \{cf. Eq.~(\ref{OsmRWM})\} we
first equilibrated the system in the $\mu VT$ ensemble, choosing  $\mu$ appropriately
to reach the desired value of ${\cal N}$. Again $3\cdot 10^5$ steps were used for this grand canonical
equilibration stage. Then the configurational bias moves were stopped, and the
${\cal N}VT$ run with the measurement of the number of monomers $N_{wall}(\lambda)$ adjacent
to the walls began.

In the configurational bias moves, one needs to utilize a biased chain
insertion method to let a polymer ``grow'' successively into the system.
At each step all possible 108 bond vectors from the current effective monomer
are examined, and a position for inserting the next monomeric unit along the
chain is chosen, respecting the excluded volume condition, and using the Boltzmann
weight calculated from the intramolecular energy, Eq.~(\ref{Estiff}). The statistical weight of
the generated polymer configuration hence is easily calculated recursively, and thus
the bias can be accounted for in the acceptance probability for the move.

In our simulations, we have recorded standard single-chain characteristics
such as the mean-square end-to-end distance $R_e$ and the mean-square gyration
radius $R_g$ of the chains ($\vec{r}_i$ are positions of monomers, $\vec{r}_{CM}$ is
the position of the center of mass),

\begin{equation}
R_e=\left<(\vec{r}_N-\vec{r}_1)^2\right>^{1/2}, \quad R_g=\left<\sum_{i=1}^N(\vec{r}_i-\vec{r}_{cm})^2/N\right>^{1/2},
\label{ReRg}
\end{equation}
from the runs in the $\mu VT$ ensemble. Note that these quantities depend on the polymer
volume fraction $\phi=8{\cal N}Na^3/V$ in the system, or the
average density of monomer units $\rho_m=\phi/8a^3$.
The average $\left<\ldots\right>$ extends over all chains and all generated system configurations.
In the SE method, where we have a density profile from a rather large density
near $z=0$ to almost zero at $z=H$, we expect that the radii $R_e, R_g$ will
depend distinctly on the height $z_{CM}$ of the center of mass of a chain, and
hence this information could only be obtained with substantially reduced
statistical accuracy. In the RWTI method, one can estimate the radii as
well if one restricts the averaging to chains with center of mass coordinate $\vec{r}_{CM}$
sufficiently remote from the walls.

\section{Results for Fully Flexible Chains}
\label{flexibleEOS}

\begin{figure}[p]
\includegraphics[width=0.7\textwidth,angle=0]{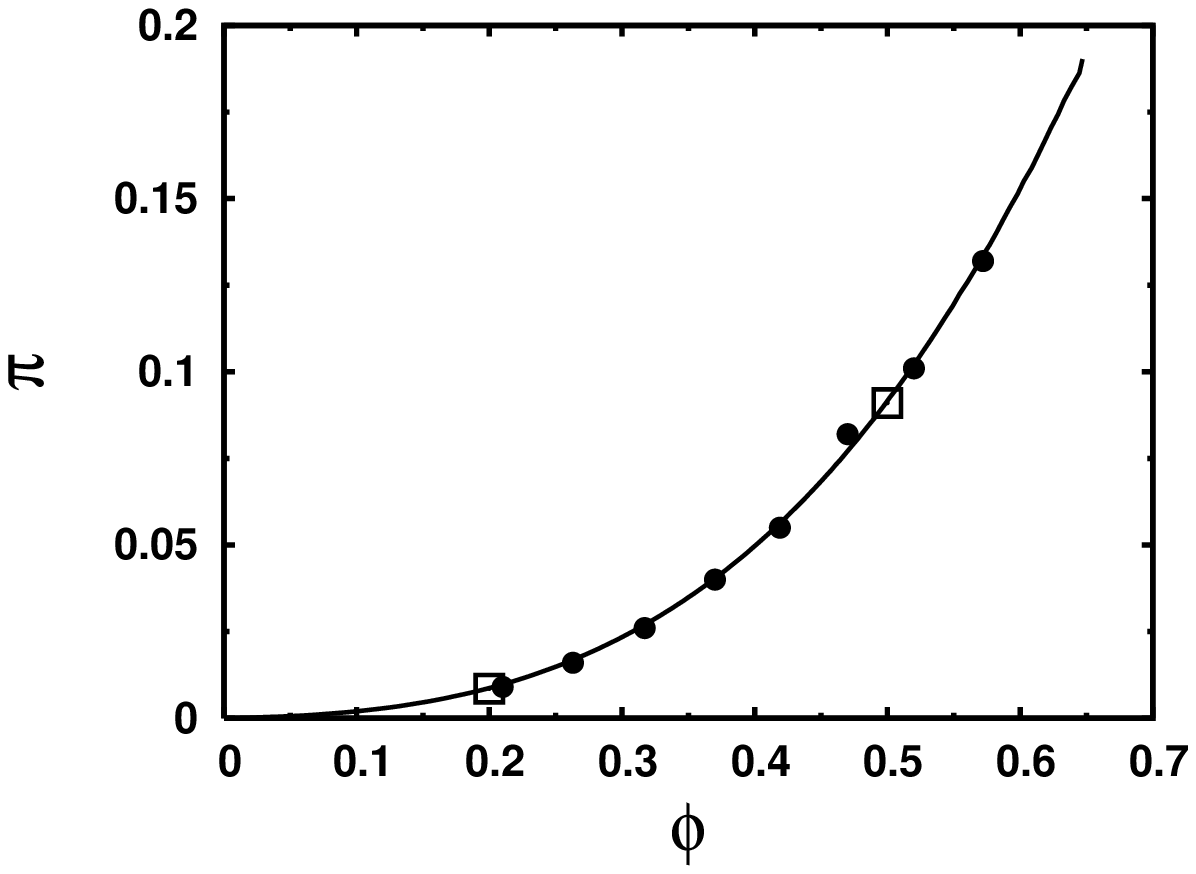}
\caption{Osmotic pressure $\pi$ plotted vs. polymer volume
fraction $\phi$, for the athermal fully flexible bond fluctuation
model on the simple cubic lattice. The solid line shows the TI$\mu
VT$ results, while the two large squares indicate results obtained
with the RWTI method. Filled circles are the corresponding RWTI
results of Deutsch and Dickman (Ref.~\onlinecite{rep_wall2}).
\label{Fig1}}
\end{figure}

We start by comparing results obtained from the RWTI method with results
obtained from the thermodynamic integration in the grand canonical ensemble (Fig.~\ref{Fig1}).
The agreement between the results of both methods actually is excellent (relative deviations
are smaller than $10^{-2}$, and the statistical error for all data points obtained by
both methods was also always less than 1\%).
Also the old data by Deutsch and Dickman~\cite{rep_wall2}, which clearly
are considerably less accurate, agree with the present calculation to within a few
percent.

An essential consideration for the TI$\mu VT$ method is that the values of $\mu$
for which calculations are performed must be so closely spaced that the probability
distributions $P_{\mu}(\rho_m)$ at neighboring choices of $\mu$ overlap strongly. This condition
has been carefully checked. Note, however, that the effort in CPU resources
to generate the full equation of state $\pi(\phi)$ in Fig.~\ref{Fig1} with the TI$\mu VT$ method
is comparable to the effort for a single point on the equation of state in the
RWTI method. Therefore we restrained our efforts to the accurate calculation
of two state points with this method only.

\begin{figure}[p]
\includegraphics[width=0.7\textwidth,angle=0]{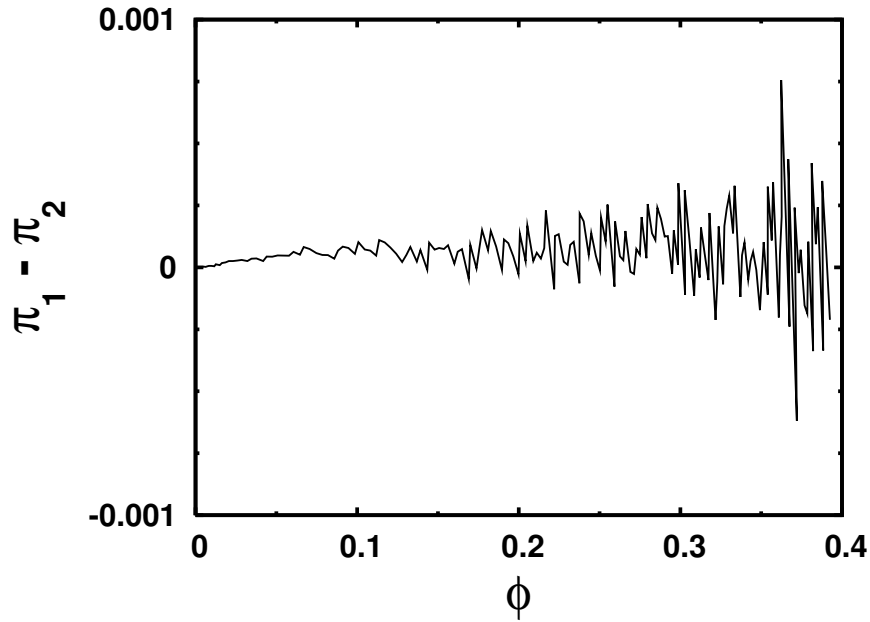}
\caption{Difference between osmotic pressure $\pi$ obtained with the
TI$\mu VT$ and SE methods plotted vs. $\phi$. Note that the
difference is less than $10^{-3}$ throughout and almost no systematic trend
can be seen (see also discussion in Section~\ref{stiffEOS} below).
The SE data were obtained for a system of size
$80 \times 80 \times 200$ containing ${\cal N}=1600$ chains, for a choice of $\lambda_g =0.01$.
\label{Fig2}}
\end{figure}

Fig.~\ref{Fig2} now compares the results obtained with the SE method with those
of the TI$\mu VT$ method, and again we find perfect agreement. Note
that the choice $\lambda_g=0.01$ implies a length $\xi_m=100$ \{Eq.~(\ref{grlength})\}, while the gyration
radius is $R_g\approx 6.4$ for $N=20$ and volume fractions in the range from $0.1<\phi<0.2$,
which are relevant here. Thus the condition $R_g\ll \xi_m$ \{Eq.~(\ref{DExtPot})\} is safely
fulfilled, but nevertheless it is important to check that no visible systematic
error of the SE method is present (to our knowledge, it is not known
in which order of $R_g/\xi_m$ systematic corrections due to the gradient in
density should be expected). As expected, we see some increase of the random
statistical error with increasing $\phi$, but the absolute magnitude of this error always stays clearly below
$10^{-3}$, and the relative error is between 1\% and 2\%.

\begin{figure}[p]
a) \includegraphics[width=0.65\textwidth,angle=0]{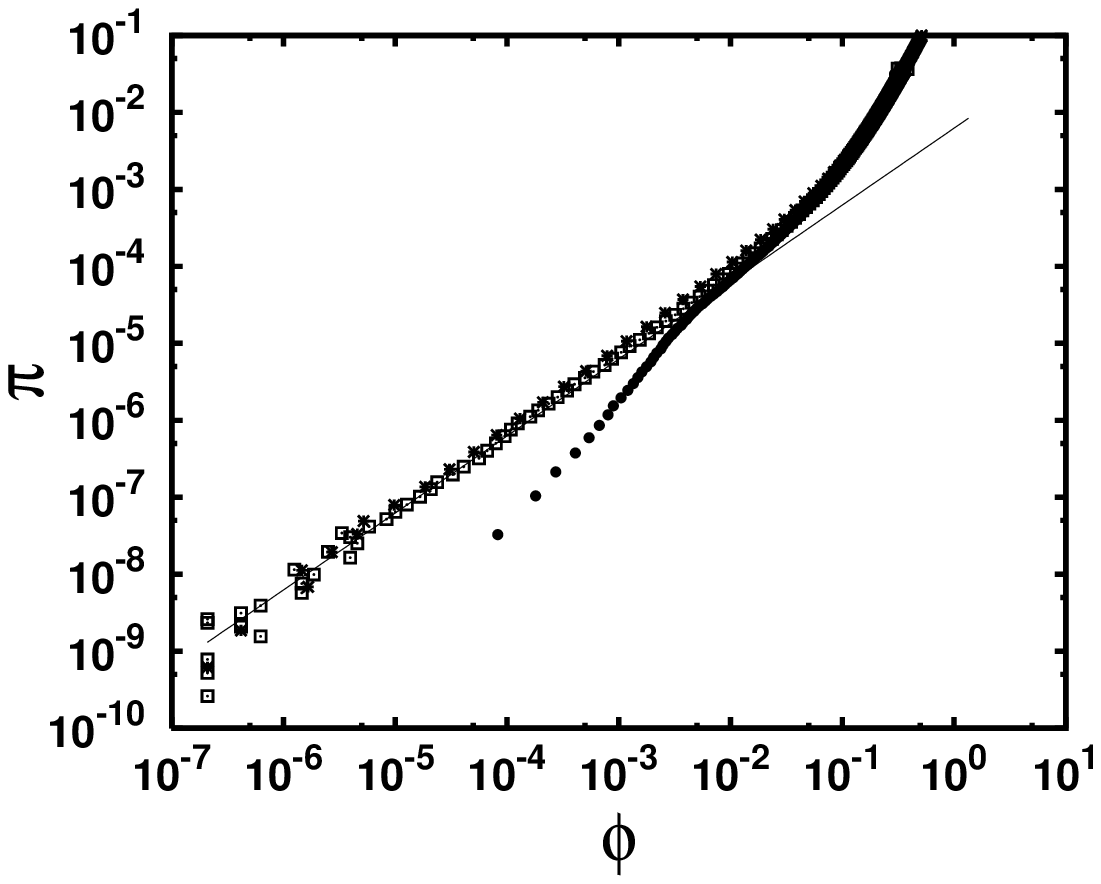}

b) \includegraphics[width=0.65\textwidth,angle=0]{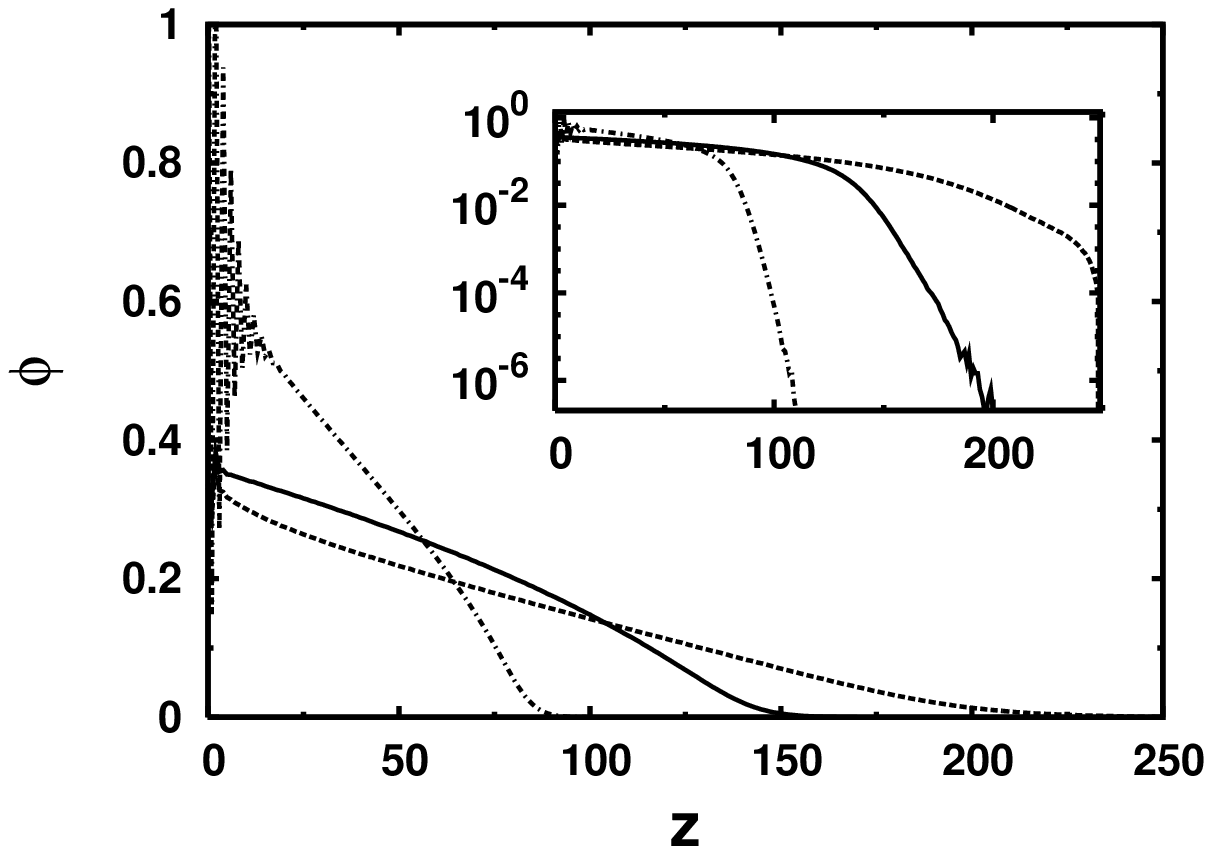}
\caption{ (a) Osmotic pressure $\pi$ (on a logarithmic scale)
plotted vs. volume fraction $\phi$ (on a logarithmic scale), for
the fully athermal solution of flexible chains of length $N=20$,
comparing several variants of the SE method for a box of size
$80\times 80 \times 250$ and ${\cal N}= 1200$. Open squares refer
to the gravitation--like potential, Eq.~(\ref{ExtPot}), with
$\lambda_g=0.01$, while the filled circles and stars refer to the
potential in Eq.~(\ref{ExtPotzk}) with $\kappa=1/2$, using
$\lambda_g^{\prime}= 0.1$  and 0.5 correspondingly. The full line is
the exactly known ideal gas limit. (b) Density
(volume fraction) profiles of monomer units,
$\phi(z) = 2 n_m(z) / n_{max} $,
corresponding to the results shown in (a);
solid line is for gravitation-like potential with $\lambda_g=0.01$,
dashed-dotted line is for $\lambda_g^{\prime}=0.5$, $\kappa=1/2$,
and dotted line is for $\lambda_g^{\prime}=0.1$, $\kappa=1/2$;
here $n_m (z)$ is the average number of monomer units at the
altitude $z$, $n_{max}=L^2/4$ is the maximal number of monomers in one
layer, and the coefficient 2 in the formulae for $\phi(z)$ above
accounts for the fact that one fully occupied layer excludes
for occupation all lattice sites in a neighboring layer.
Note that the density profile for the case
$\lambda_g^{\prime}= 0.1$ is affected by the wall at $z=250$
(the inset shows the same figure using a logarithmic scale for the density),
leading to systematic errors in the equation of state. The decay
of density profiles always occurs on scales much larger than the
average value of gyration radius $\sqrt{\langle R_g^2 \rangle}$
which was about $6.4$ for dilute region and about $5.6$ in the
concentrated region close to the wall. \label{Fig3}}
\end{figure}

\begin{figure}[p]
a) \includegraphics[width=0.65\textwidth,angle=0]{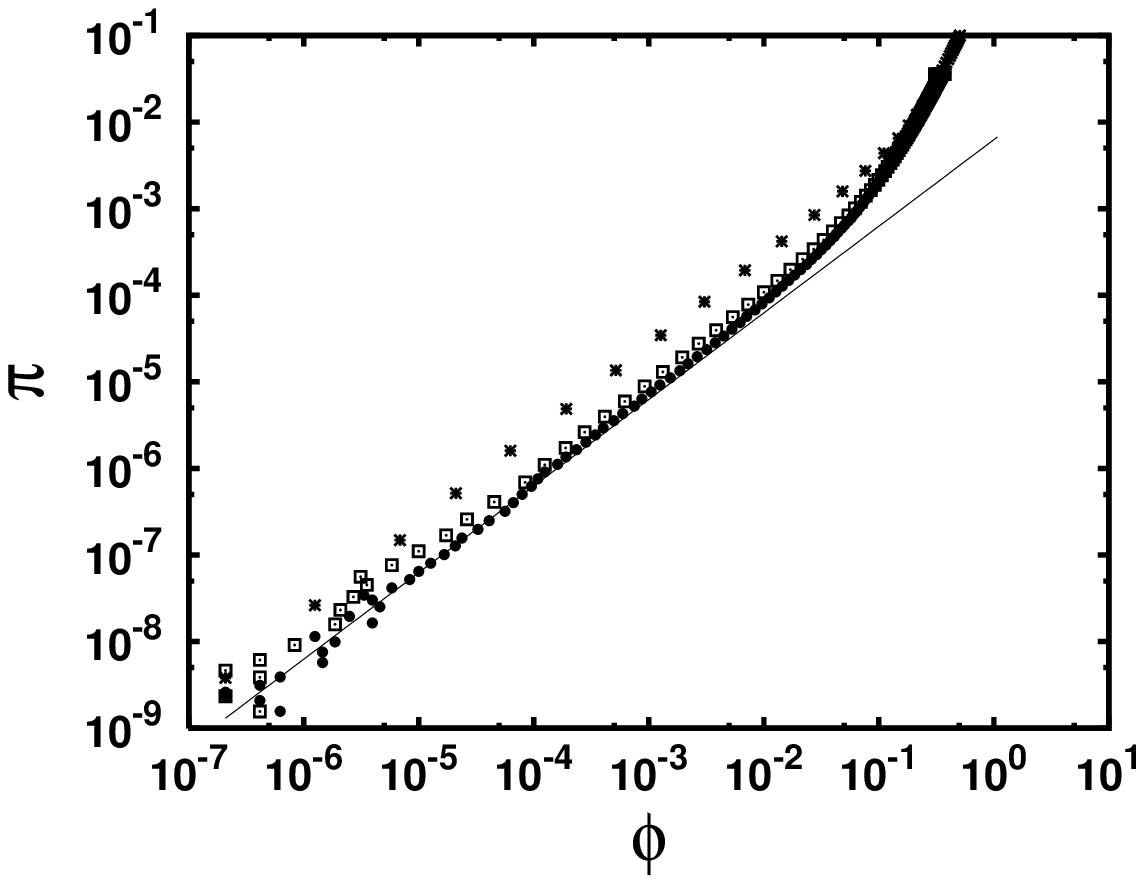}

b) \includegraphics[width=0.65\textwidth,angle=0]{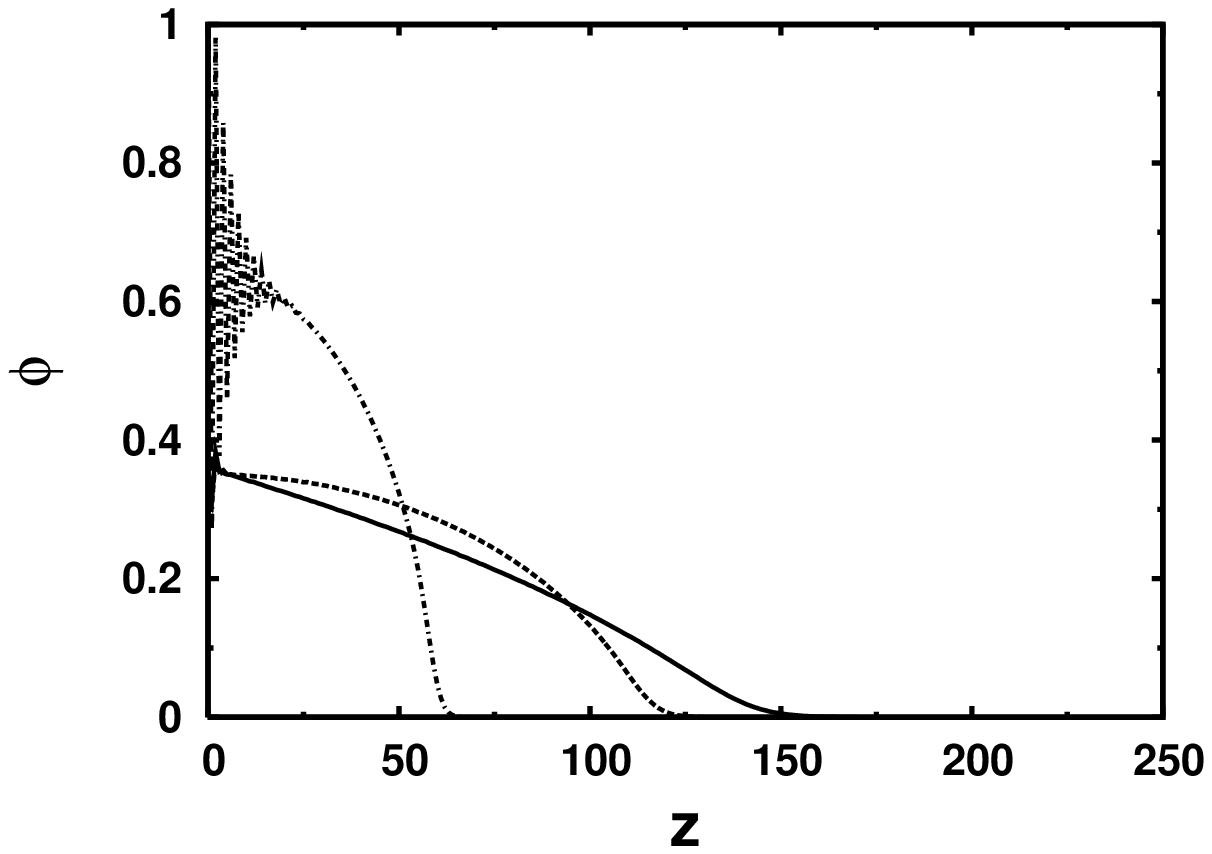}
\caption{(a) Same as Fig.~\ref{Fig3}a, but comparing the results
for the gravitation-like potential, Eq.~(\ref{ExtPot}), with
$\lambda_g= 0.01$ (filled circles) to results for the potential in
Eq.~(\ref{ExtPotzk}) with $\kappa=2$, using $\lambda_g^{\prime}=
0.0001$ (open squares) and $\lambda_g^{\prime}= 0.001$ (stars).
(b) Density (volume fraction) profiles of monomer units
corresponding to the results shown in Fig.~\ref{Fig4}a;
solid line is for gravitation-like potential with $\lambda_g=0.01$,
dashed-dotted line is for $\lambda_g^{\prime}=0.001$, $\kappa=2$,
and dotted line is for $\lambda_g^{\prime}=0.0001$, $\kappa=2$.
\label{Fig4}}
\end{figure}

Thus we confirm the conclusion of Addison et al.~\cite{Hansen}, for a different model
than studied in their work, that the SE method can yield a reliable estimation of the
equation of state. However, care has to be exerted that parameters such as the
height $H$ of the simulation box and the strength of the external potential $\lambda_g$ are chosen
appropriately, and also the statistical effort needs to be large enough.
Figs.~\ref{Fig3} and~\ref{Fig4}
contain examples of the problems that one encounters when these conditions
are not met. E.g., when the potential is chosen too weak for the chosen size $H$ (and
number of chains ${\cal N}$), so that the density profile (we plot here and below
profiles of the polymer volume fraction $\phi$) is slightly affected by the hard
wall at $z=H$ (see Fig.~\ref{Fig3}b), a systematic depression of $\pi(\phi)$
in the ideal gas region (where simply $\pi=\rho=\rho_m/N=\phi/8Na^3$ must hold)
is found (see Fig.~\ref{Fig3}a). Conversely, when the statistics does not suffice, or equilibration time was
too short so that the asymptotic behavior of the density profile (see Eq.~\ref{barom} below)
has not been reached, one can also get an overestimation of the pressure in this region.
Interestingly, even such data that are invalid in the ideal gas region still
merge rather well at larger volume fractions, indicating the robustness of the
SE method in this regime. It is also remarkable that the strong layering
found for the potential proportional to $z^{1/2}$ near the wall for $\lambda^{\prime}_g=0.5$
(and the subsequent rather rapid decrease of the density towards zero) do
not create any problems, however. Fig.~\ref{Fig4}, on the other hand, shows a case
($\kappa=2$ in the potential Eq.~\ref{ExtPotzk}, $\lambda^{\prime}_g=0.001$)
for which the resulting density profile (Fig.~\ref{Fig4}b) is too steeply varying, and
then it is very likely that the local density approximation in no longer accurate.
Consequently, it is no surprise that the osmotic pressure comes out systematically
too large (Fig.~\ref{Fig4}a). However, the choice $\kappa=2$, $\lambda^{\prime}_g=0.0001$
is again in very good agreement with the result obtained for the potential Eq.~(\ref{ExtPot})
with $\lambda_g=0.01$ (which also agrees with the TI$\mu VT$ results, as noted above).
The finding that the three potentials, which lead to very different density profiles,
nevertheless yield the same result for the equation of state $\pi(\phi)$ is a very clear
evidence that the local density approximation is valid, for the chosen set of parameters.

\begin{figure}[p]
\includegraphics[width=0.7\textwidth,angle=0]{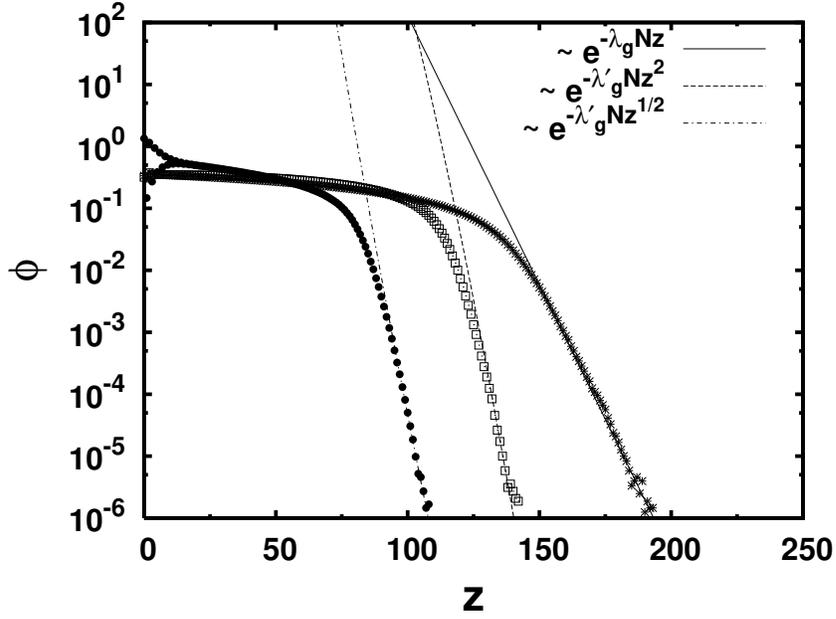}
\caption{Semi-log plot of the density profiles of
monomer units for three choices of the external potential:
Eq.~(\ref{ExtPot}) with $\lambda_g= 0.01$ (stars) with the
asymptotic barometric law (solid line), Eq.~(\ref{ExtPotzk}) with
$\kappa=1/2$, $\lambda_g^{\prime}= 0.5$ (filled circles) with the
corresponding asymptotic law (dashed-dotted line) and with
$\kappa=2$, $\lambda_g^{\prime}= 0.0001$ (open squares) with the
asymptotic law (dashed line). For all cases the box size was $80
\times 80 \times 250$ and the number of chains was ${\cal N}=
1200$. \label{Fig5}}
\end{figure}

Fig.~\ref{Fig5} presents then a comparison of the monomer density profiles for the cases
where the SE method yields correct results. We use here a logarithmic scale to demonstrate
that for low densities the barometric height formula
\begin{equation}
\rho_m(z)\propto \exp[-U_{\rm external}(z)/k_BT]
\label{barom}
\end{equation}
is fulfilled. One can see that the density profiles in Fig.~\ref{Fig5} are compatible with Eq.~(\ref{barom})
over several decades
in density. Only for extremely small densities the statistical scatter becomes significant.
Thus, analyzing the data for $\rho_m(z)$ in this way provides a check whether sufficient
statistical effort has been invested.

\section{Variable Chain Stiffness}
\label{stiffEOS}

Allowing for nonzero parameters $f$ and $\varepsilon_0$ in Eq.~(\ref{Estiff}) already has interesting effects
even in the ideal gas limit where single chain properties dominate the behavior, and
this we will discuss first.

\begin{figure}[p]
a) \includegraphics[width=0.65\textwidth,angle=0]{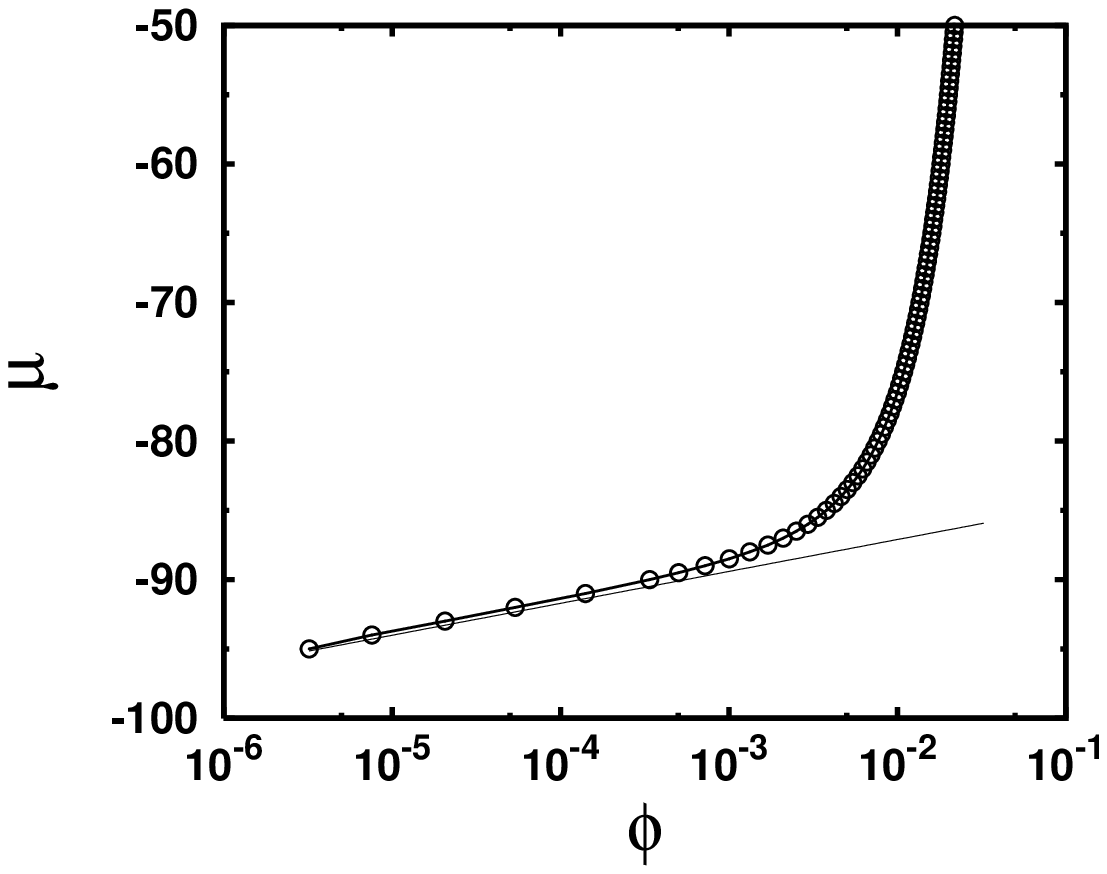}

b) \includegraphics[width=0.65\textwidth,angle=0]{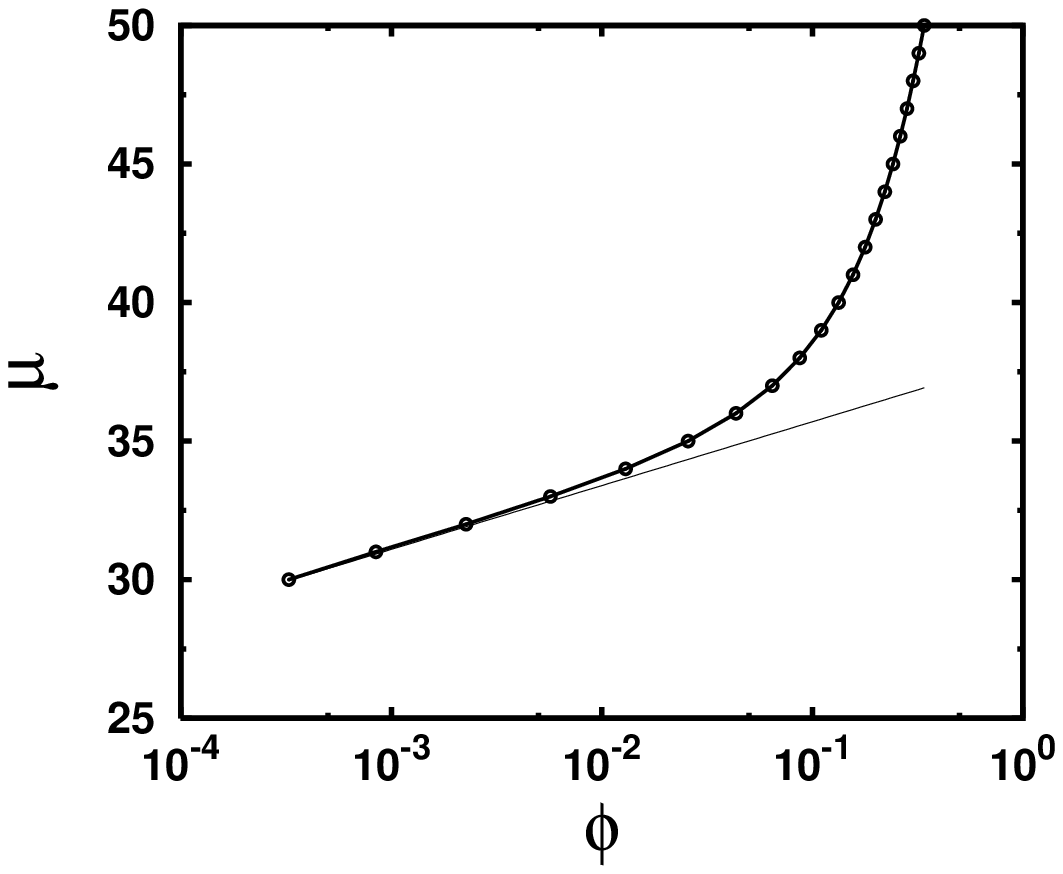}
\caption{Chemical potential per chain vs. volume fraction $\phi$,
for the case of fully flexible chains, (a), where both $f=0$ and
$\varepsilon_0 = 0$, and for semiflexible ones, (b), with
$f=2.68$, $\varepsilon_0 = 4$. Note the logarithmic scale of the
abscissa. Straight solid line indicates the fit to
Eq.~(\ref{muideal}), with $C(0,0)=-82.5$ and $C(2.68,4)=38.0$,
respectively. \label{Fig6}}
\end{figure}

Fig.~\ref{Fig6} shows the chemical potential per chain, $\mu(\phi)$, as function of the volume
fraction $\phi$ both for fully flexible and for semiflexible chains. The logarithmic
scale of the abscissa is used to demonstrate the approach to the ideal gas law,
\begin{equation}
\mu_{id}(\phi)=\ln\phi + C(f,\varepsilon_0),
\label{muideal}
\end{equation}
where $C(f,\varepsilon_0)$ is a constant that depends on both $f$ and $\varepsilon_0$. The data shown
in Fig.~\ref{Fig6} were obtained from simulations in the $\mu VT$ ensemble, of course.

\begin{figure}[p]
a) \includegraphics[width=0.65\textwidth,angle=0]{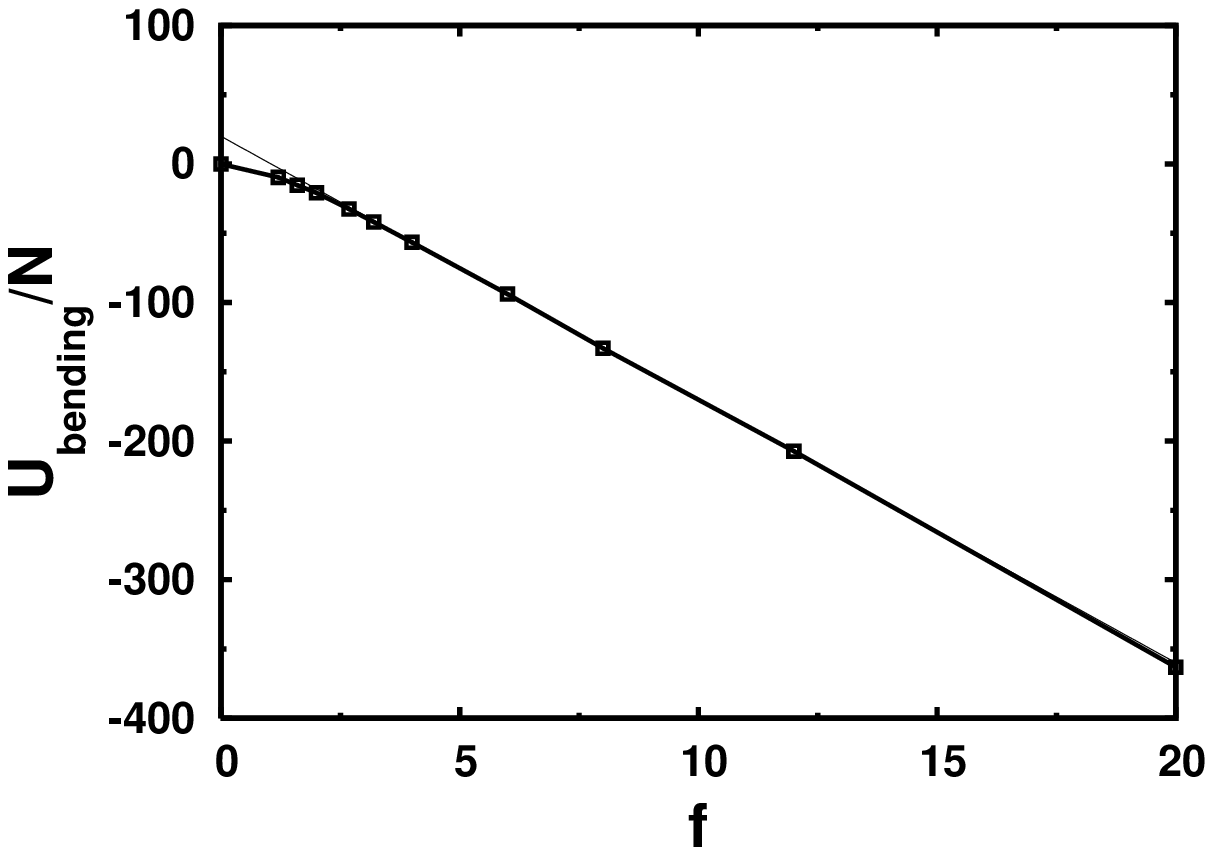}

b) \includegraphics[width=0.65\textwidth,angle=0]{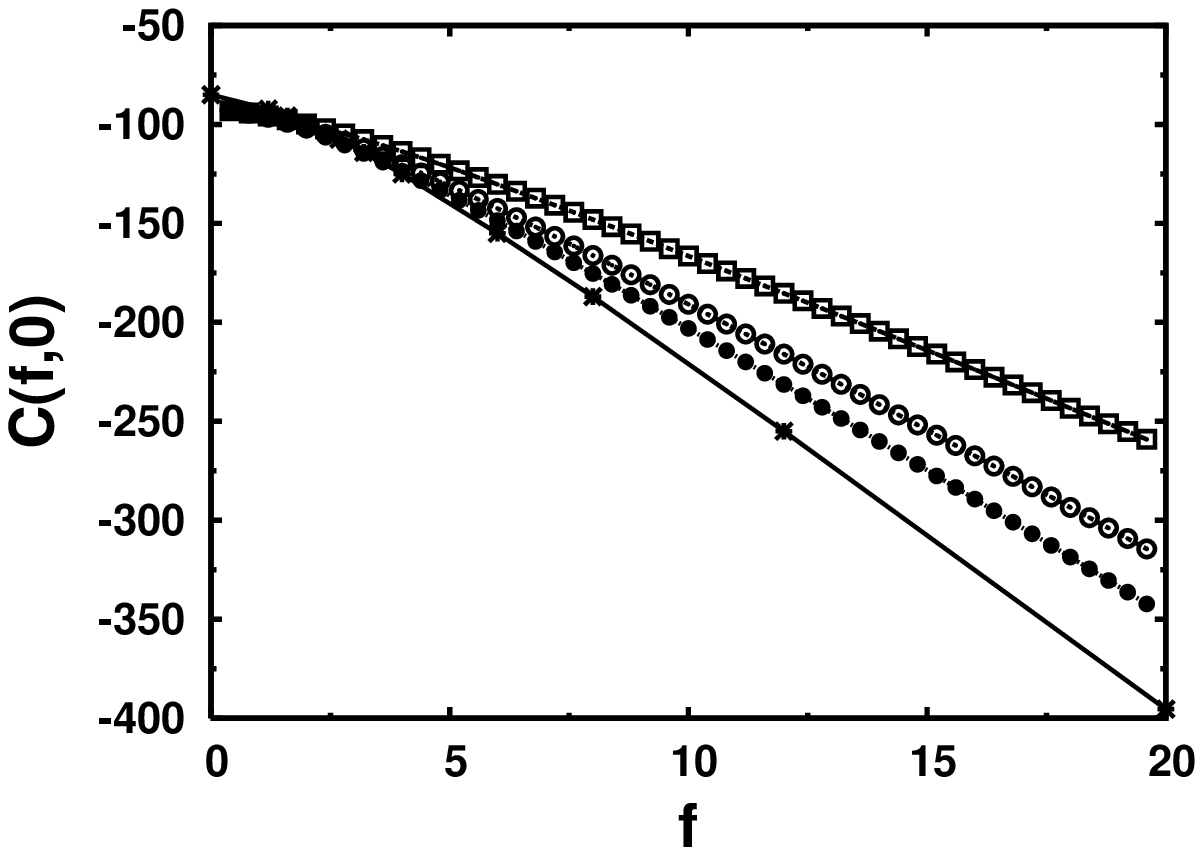}
\caption{ Bending energy per chain (a) and $C(f,0)$ (b) plotted
vs. the stiffness parameter $f$. Thin solid line in (a) indicates the linear
fit. In part (b) $C(f,0)$ is plotted with stars ($*$) and solid line, while symbols
indicate approximations where $C(f,0)$ is found from a
decomposition of the chain partition function in terms of
independent trimers (open squares), tetramers (open circles) and
pentamers (filled circles).
\label{Fig7}}
\end{figure}

\begin{figure}[p]
a) \includegraphics[width=0.65\textwidth,angle=0]{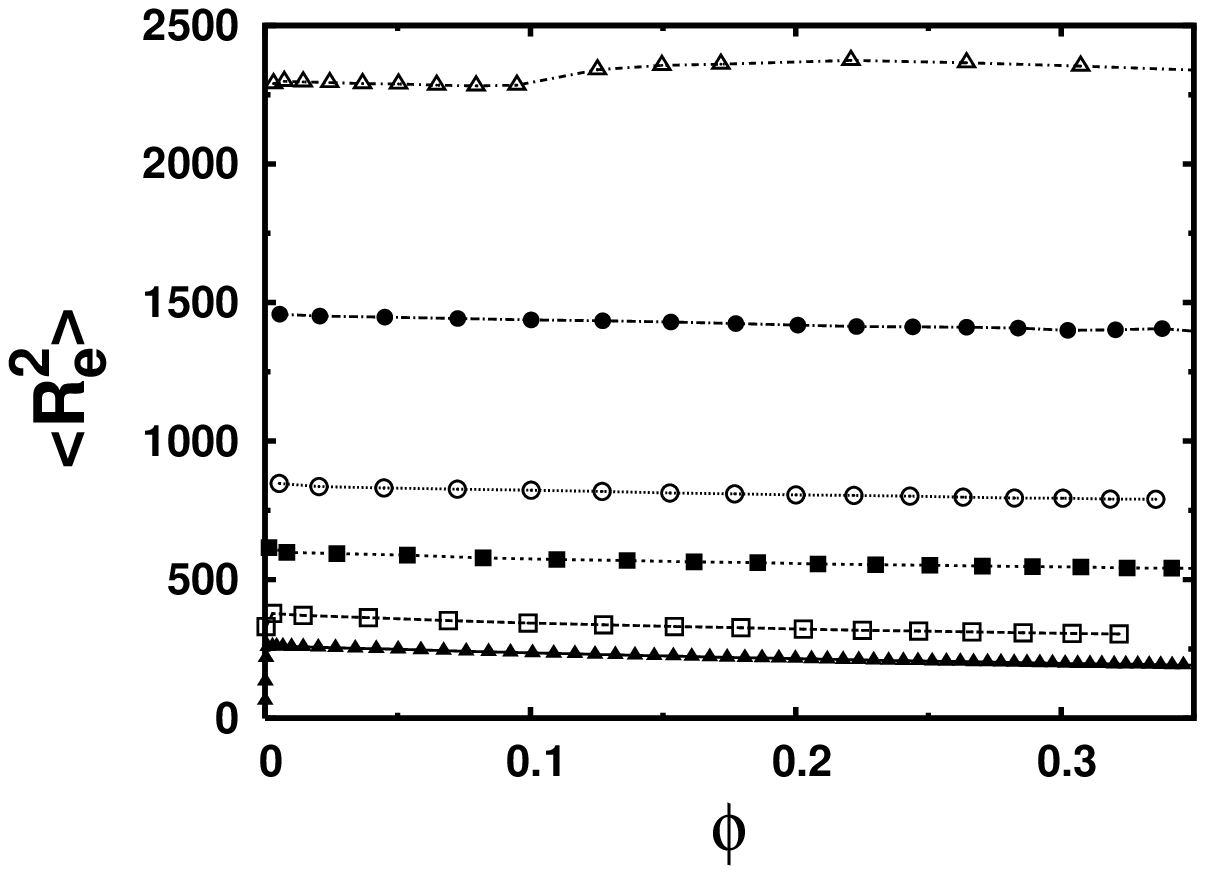}

b) \includegraphics[width=0.65\textwidth,angle=0]{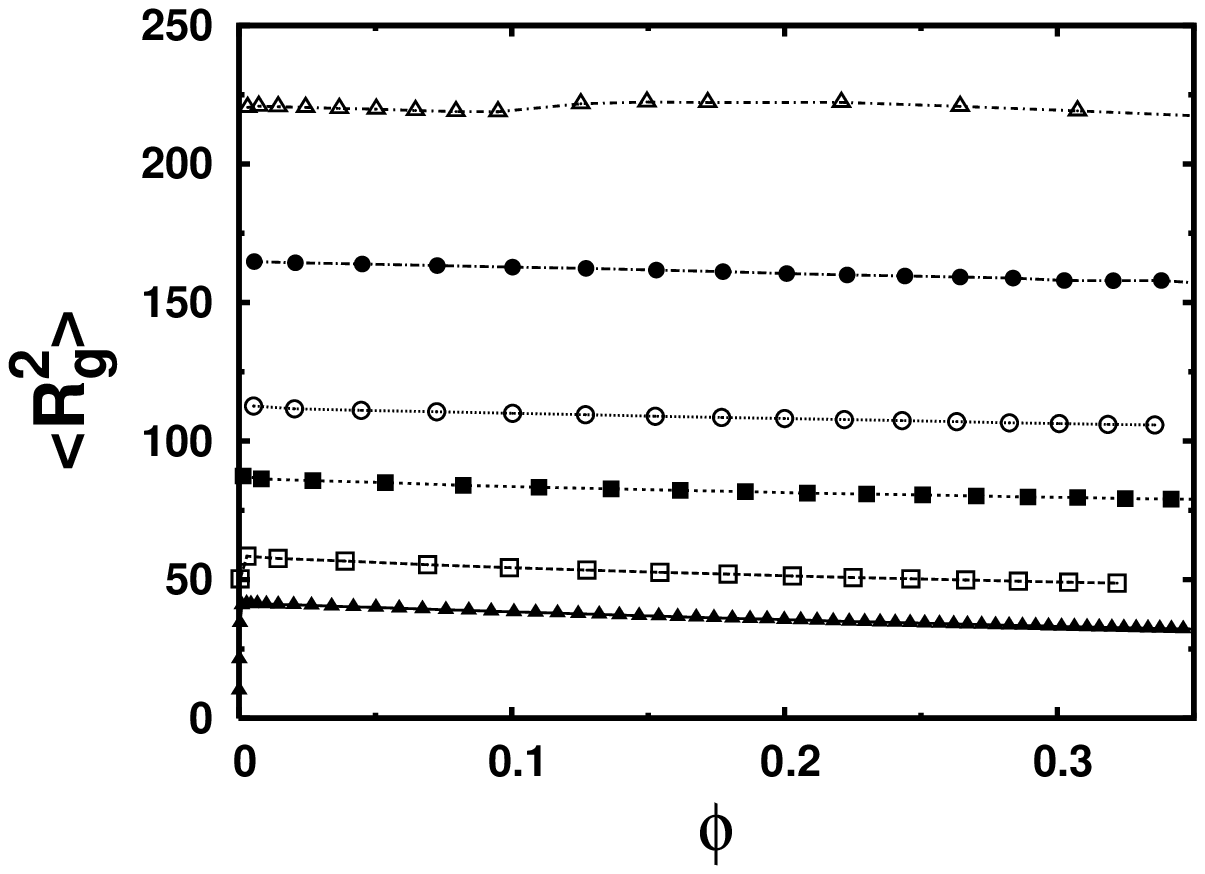}
\caption{Mean squared end-to-end distance (a) and mean squared
gyration radius (b) plotted vs. volume fraction $\phi$, for
several choices of the stiffness parameter $f$, $f=0$ (filled
triangles), $f=1.2$ (open squares), $f=2.68$ (filled squares),
$f=4$ (open circles), $f=8$ (filled circles), $f=20$ (open
triangles), and $\varepsilon_0=0$. \label{Fig8}}
\end{figure}

Fig.~\ref{Fig7} shows the dependence of both the -bending energy, $U_{\rm bending}$ \{Eq.~\ref{Estiff} \},
and of $C(f,0)$ on the stiffness parameter $f$. One sees that the bending energy per
chain is essentially decreasing linearly with $f$ for $f\ge 5$, and then also the bending
energy per monomer clearly exceeds the thermal energy, and hence the chains get
strongly stretched (Fig.~\ref{Fig8}). The linear variation of the bending energy with $f$ in Fig.~\ref{Fig7}a
indicates that the bending energy is already fully ``saturated'', i.e. an energetic
minimum is reached. An asymptotic formula for the bending energy per chain for the case of fully stretched chain
gives $U_{\rm bending}/{\cal N}=-1.03 (N-2) f$, i.e. for $f=20$, $N=20$ the bending energy per chain is equal
to $370.8$ (cf. Fig.~\ref{Fig7}a).

The variation of the constant $C(f,\varepsilon_0)$ with $f$ does not have an immediately
obvious interpretation, however. In order to interpret this behavior analytically,
we have estimated $C(f,\varepsilon_0)$ applying approximate single chain partition functions
in terms of the independent trimer, quadrumer and pentamer approximations~\cite{trimer1,trimer2}.
One can see (Fig.~\ref{Fig7}b) that these approximations follow the same trend as our fit to the
numerical data, but clearly the convergence of these approximations to the numerical
result from the simulation is rather slow. Obviously, the constant $C(f,0)$ reflects
a delicate interplay between excluded volume and chain stiffness contributions
to the single chain entropy that this constant measures.

\begin{figure}[p]
a) \includegraphics[width=0.65\textwidth,angle=0]{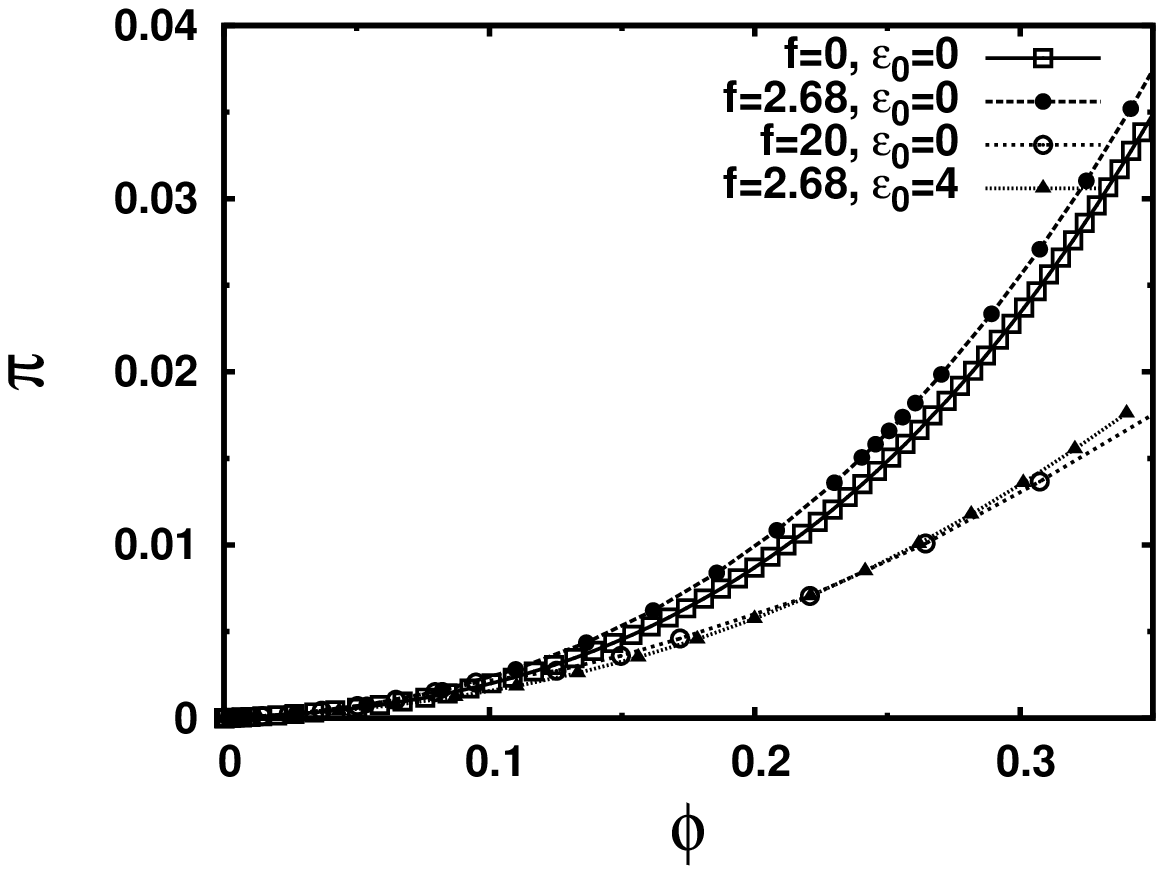}

b) \includegraphics[width=0.65\textwidth,angle=0]{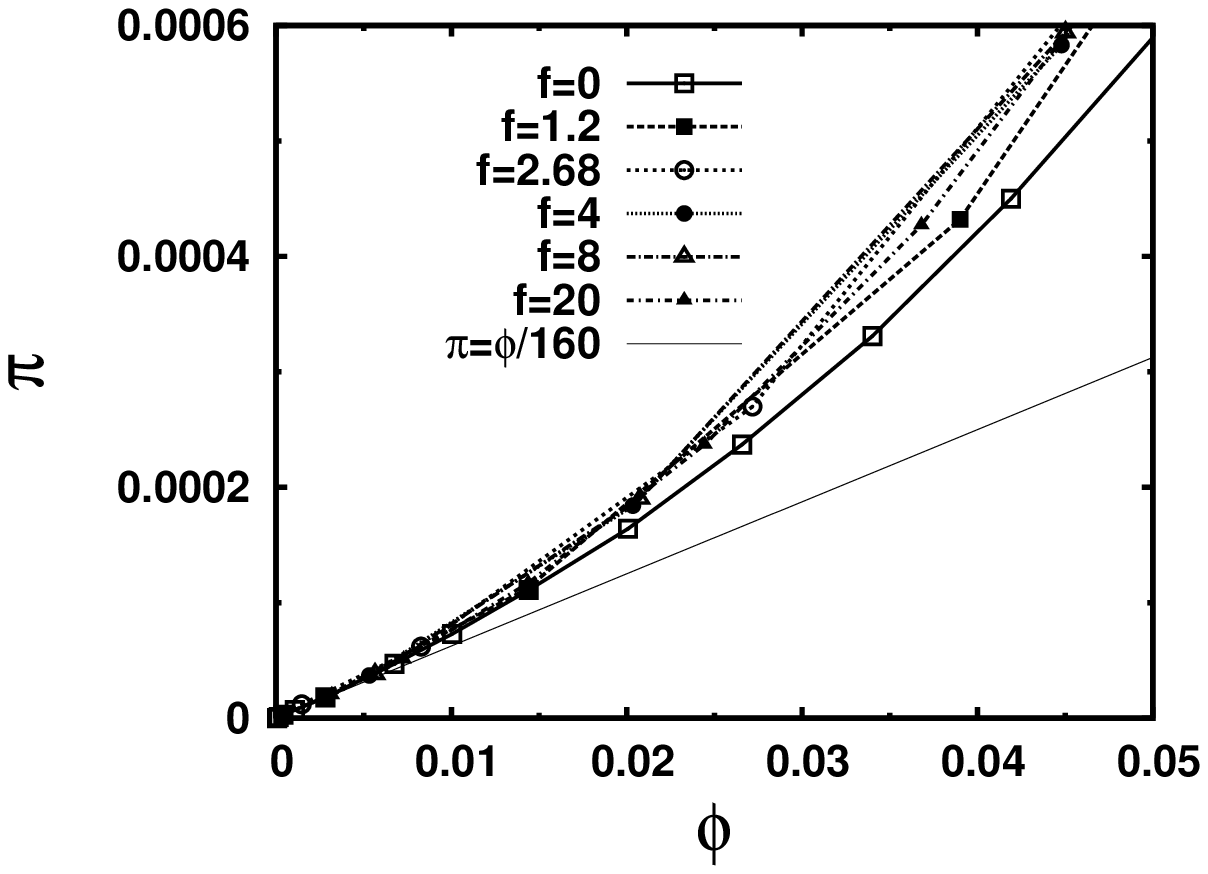}
\caption{Equation of state $\pi(\phi)$ plotted vs. $\phi$, for
several choices of the stiffness parameter $f$ and two choices for
the bond length energy parameter $\varepsilon_0$ (a). Magnified
view of the equation of state in the low density regime (b); the ideal
gas limit $\pi = \phi / 160$ is indicated by a thin solid line. The
choices of the parameter $f$ (and $\varepsilon_0$ in part (a);
part (b) refers to $\varepsilon_0=0$ only) are given in the
figure. Curves are obtained using the TI$\mu VT$ method.
\label{Fig9}}
\end{figure}

Knowing $\mu_{id}(\phi)$ we obtain $\mu_{ex}(\phi)=\mu-\mu_{id}(\phi)$, which is needed for the
thermodynamic integration to obtain the pressure \{Eqs.~(\ref{press})--(\ref{pii})\}. The result is presented
in Fig.~\ref{Fig9}. One sees that the variation of chain stiffness at not too large volume
fractions has surprisingly little effect on the equation of state, despite the strong
change in chain extensions (Fig.~\ref{Fig8}) and structure, and the decrease in the entropy
of the chains (Fig.~\ref{Fig7}). We note that the osmotic pressure for semiflexible chains gets
slightly enhanced, in comparison  to the fully flexible ones, as soon as one reaches
a volume fraction of about $\phi\approx 0.01$, where the first deviations from the ideal gas
law $\pi=\phi/8Na^3=\phi/160$ occur (Fig.~\ref{Fig9}b). For large $\phi$, however, only the data for not so
large $f$ lie above the curve for $f=0$, while e.g. the data for $f=20$ lie below those
for  $f=0$ if $\phi>0.12$. This implies that at large enough $\phi$ the variation of
$\pi$ with $f$ is non-monotonic: $\pi$ increases first, and then decreases again. Presumably
this decrease reflects the onset of a nematic short range order --- small clusters of stretched chains
oriented more or less in parallel take less free volume than randomly oriented
ones, and hence lead to a decrease of osmotic pressure.

\begin{figure}[p]
a) \includegraphics[width=0.65\textwidth,angle=0]{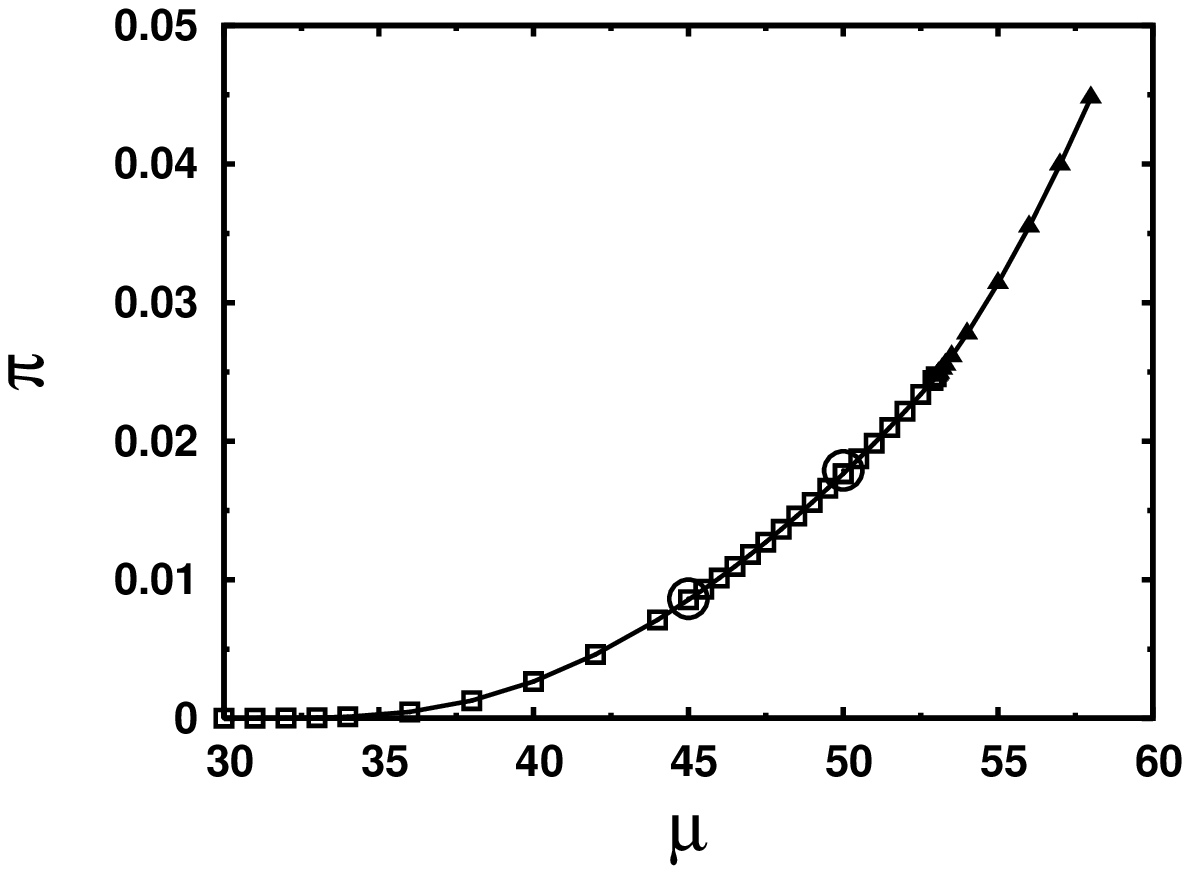}

b) \includegraphics[width=0.65\textwidth,angle=0]{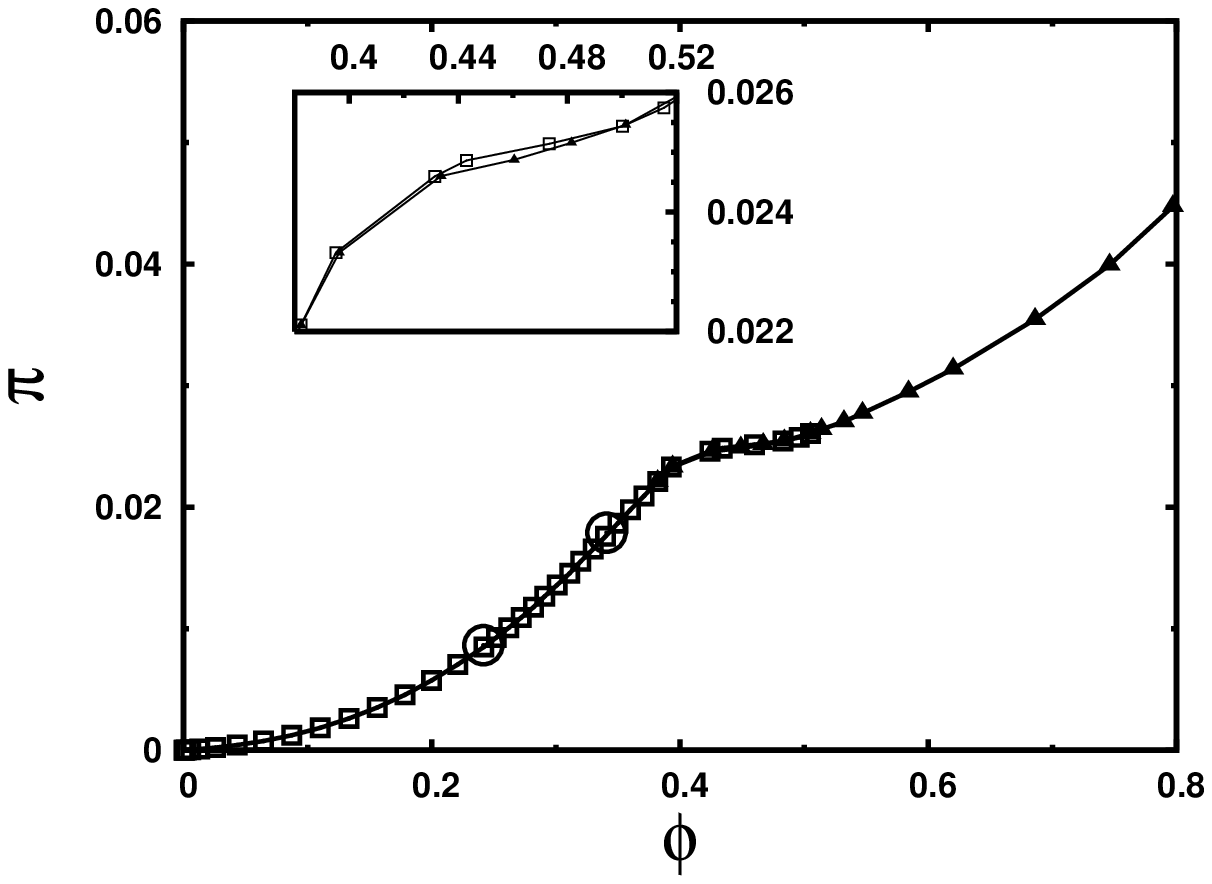}
\caption{Osmotic pressure $\pi$ plotted vs. the chemical potential
$\mu$ (a) and versus the volume fraction $\phi$ (b), for the model
with parameters $f=2.68$ and $\varepsilon_0=4$. Curves are
obtained using the TI$\mu VT$ method (filled triangles correspond
to a dense packed starting conformation, while open squares to a
dilute isotropic one), the two large open circles show results obtained
from the RWTI method. The inset shows an enlarged region in the
vicinity of the isotropic--nematic transition. \label{Fig10}}
\end{figure}

Of course, this interpretation of the pressure maximum as function of $f$ is
highly speculative, and a clear answer must await a careful analysis of the
isotropic-nematic transition in this model.
Figure~\ref{Fig10} shows that our techniques are suitable to locate the transition.
Here we first consider the case $f=2.68$, $\varepsilon_0=4$, which was studied in
our previous work on the isotropic--nematic transition for this model~\cite{Weber,LCpaper1}.
Although one expects this transition to be weakly of first order, and hence the $\pi(\phi)$ curve should
have a small two-phase coexistence region, it turns out that for these values of parameters
the width of this two-phase coexistence region is unmeasurably
small, on the scale of Fig.~\ref{Fig10} it cannot be resolved. Thus, the isotropic and nematic
branches $\pi_{iso}(\mu)$, $\pi_{nem}(\mu)$ in Fig.~\ref{Fig10} meet at the transition point in the diagram
almost tangentially, and in the $\pi(\phi)$ curve the transition shows up only as a slightly
rounded kink. This behavior is compatible with previous studies of this model~\cite{Weber,LCpaper1}.
This lack of a two-phase region at the transition allows to carry out the integration
in Eqs.~(\ref{press})--(\ref{pii}) from the isotropic phase over the transition point into the region of the
nematic phase (so no reference state in the nematic phase is required).

\begin{figure}[p]
\includegraphics[width=0.7\textwidth,angle=0]{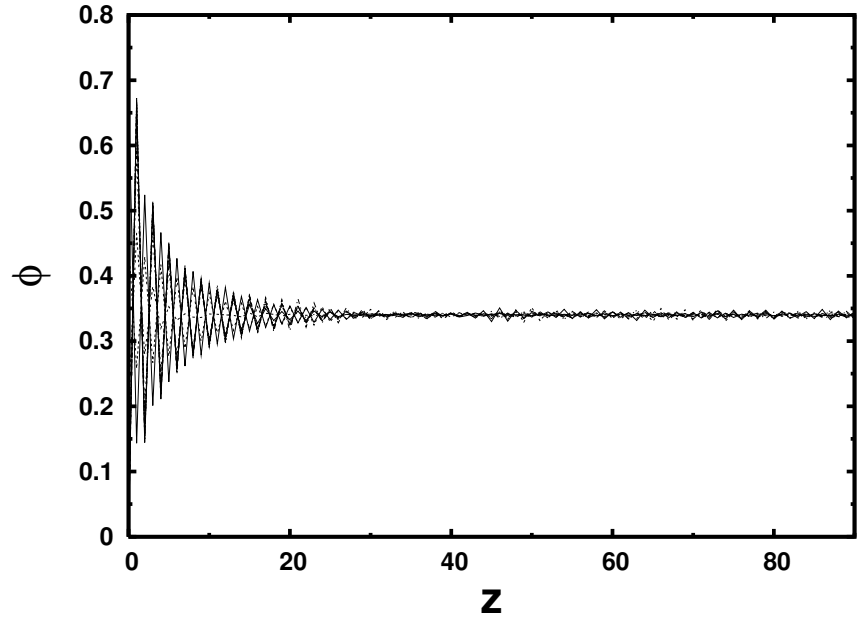}
\caption{Density (volume fraction) profile
of a $60\times 60 \times 180$ system (only half of the symmetric
profile is shown) for the parameters $f=2.68$ and
$\varepsilon_0=4$ at $\phi \approx 0.34$, as used for the calculation of the pressure $\pi(\phi)$ in the
RWTI method. Different curves for different values of the repulsion
parameter $\lambda$ from $\lambda=0.02$ to $\lambda = 1.0$ are superimposed.
\label{Fig11}}
\end{figure}

Applying the RWTI method, a very pronounced layering (extending over
about 20 lattice units) was observed (Fig.~\ref{Fig11}) for the state point with the higher volume
fraction ($\phi \approx 0.34$) in Fig.~\ref{Fig10}. Close to the repulsive walls, orientational ordering
(or even almost crystalline packing) was observed. However, in the center of the system
a homogeneous state at bulk density was clearly reached (Fig.~\ref{Fig11}), and the
agreement of the pressure estimates obtained (Fig.~\ref{Fig10}) suggests that the observed
layering (Fig.~\ref{Fig11}) does not invalidate the RWTI method here.

\begin{figure}[p]
a) \includegraphics[width=0.55\textwidth,angle=0]{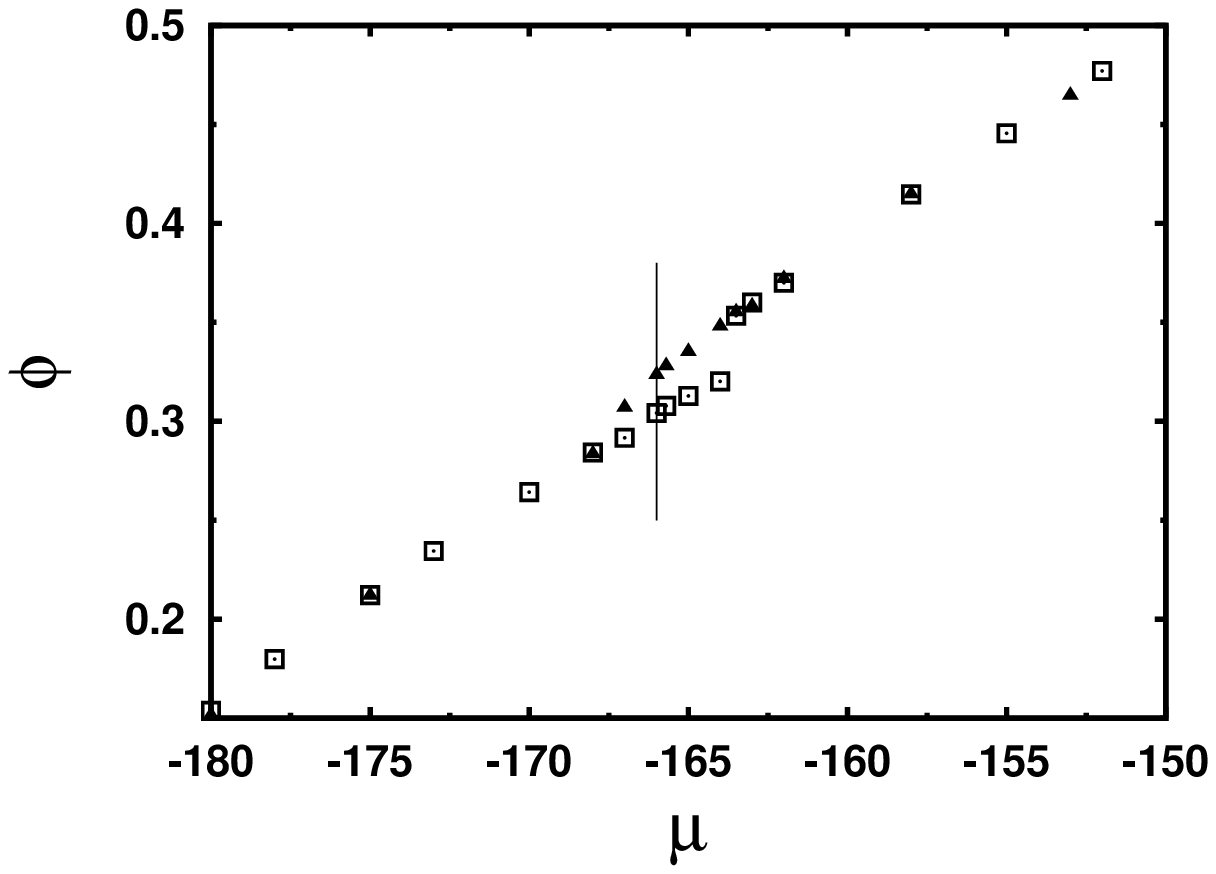}

b) \includegraphics[width=0.55\textwidth,angle=0]{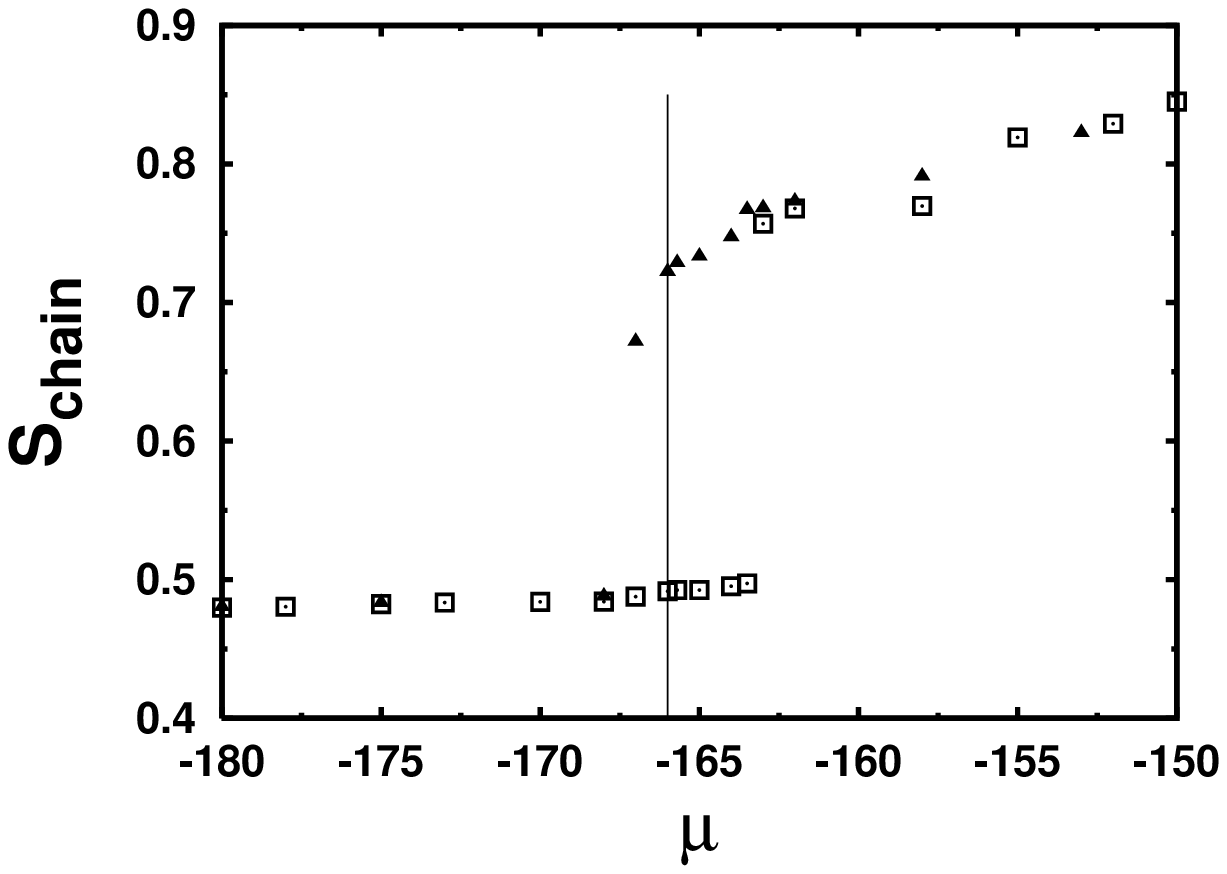}

c) \includegraphics[width=0.55\textwidth,angle=0]{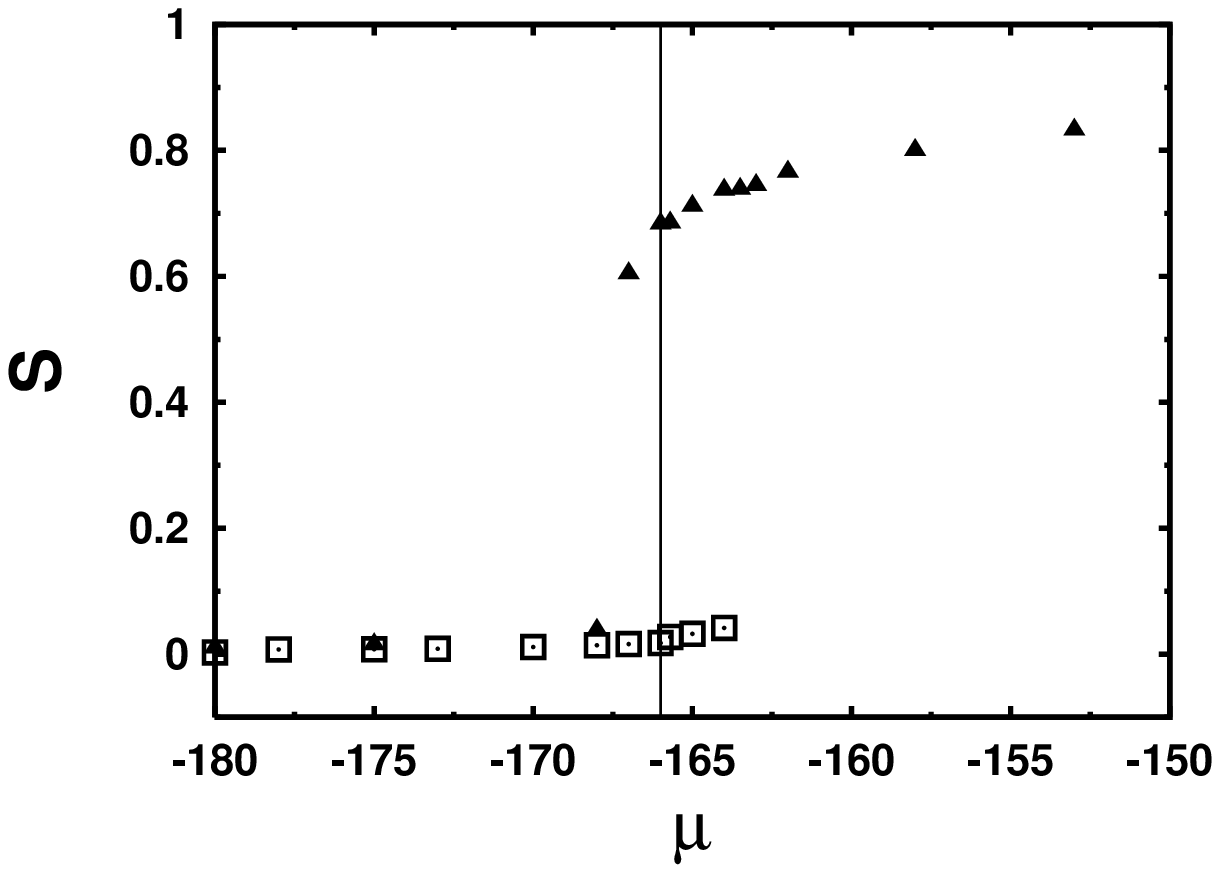}
\caption{Hysteresis for the density vs. $\mu$ dependence (a), for
the dependence of the single-chain orientational order parameter, $S_{chain}$,
vs. $\mu$ (b) and the total orientational order parameter, $S$, vs. $\mu$ (c).
Filled triangles correspond to a dense packed
starting conformation, while open squares correspond to a dilute
isotropic starting conformation. In part (c) for the case of a
dilute isotropic starting conformation only data points for $\mu$
below the jump to a nematic state are presented.
Vertical lines show the estimated value
of the chemical potential at the transition point, $\mu = -166 \pm 0.5$
(determined in Fig.~\ref{Fig16}).
\label{Fig13}}
\end{figure}

Let us now consider the case of stiffer macromolecules, $f=8$, $\varepsilon_0=0$.
We performed grand canonical simulations in the cubic box $L = 90$, $H=90$ (box 1)
and elongated box $L = 80$, $H = 150$ (box 2).
Two starting conformations have been used --- the completely empty box and the maximally dense
packed box with chains placed along one coordinate axis having bond lengths equal to $2$.
The difference between these boxes with different geometries and sizes is that it is possible to
fill the second one (box 2) with a volume fraction equal to $1$, while this is impossible for the
first box (box 1).
To characterize the orientational ordering of
the bonds we have estimated the standard $3\times 3$ nematic order parameter tensor
\begin{equation}
Q_{\alpha \beta }=\frac{1}{{\cal N} (N-1)}\sum\limits_{i=1}^{{\cal N} (N-1)}\frac{1}{2}
(3 e_{i}^{\alpha }e_{i}^{\beta } -\delta _{\alpha \beta }),
\label{Qtensor}
\end{equation}
where $e_{i}^{\alpha }$ is the $\alpha $-th component of the unit vector
along the bond connecting monomers $i$ and $i+1$ (the largest eigenvalue
of this tensor is the nematic orientational order parameter $S$).
In order to distinguish between different types of nematic structures
(e.g., monodomain vs. multidomain structures) observed in the simulation~\cite{LCpaper1}
we have also calculated the largest eigenvalue of this tensor for each chain separately,
and afterwards performed the averaging over all chains in the system
(the single-chain orientational order parameter obtained in such a way is denoted $S_{\rm chain}$).
The hysteresis for the dependencies of the density (volume fraction) $\phi$,
the total orientational order parameter $S$
and of the single-chain orientational order parameter $S_{\rm chain}$
vs. the chemical potential for simulations in box 2 is shown in Fig.~\ref{Fig13}.

\begin{figure}[p]
a) \includegraphics[width=0.65\textwidth,angle=0]{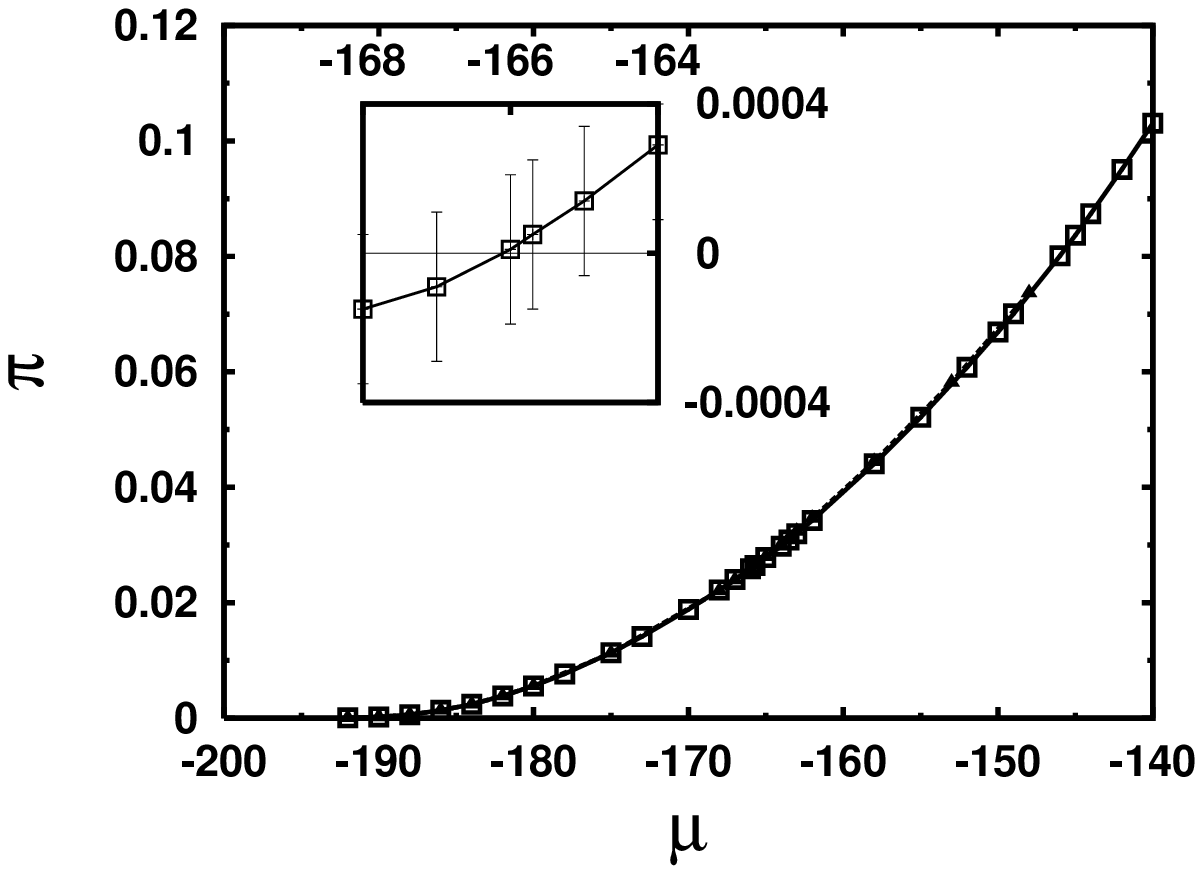}

b) \includegraphics[width=0.65\textwidth,angle=0]{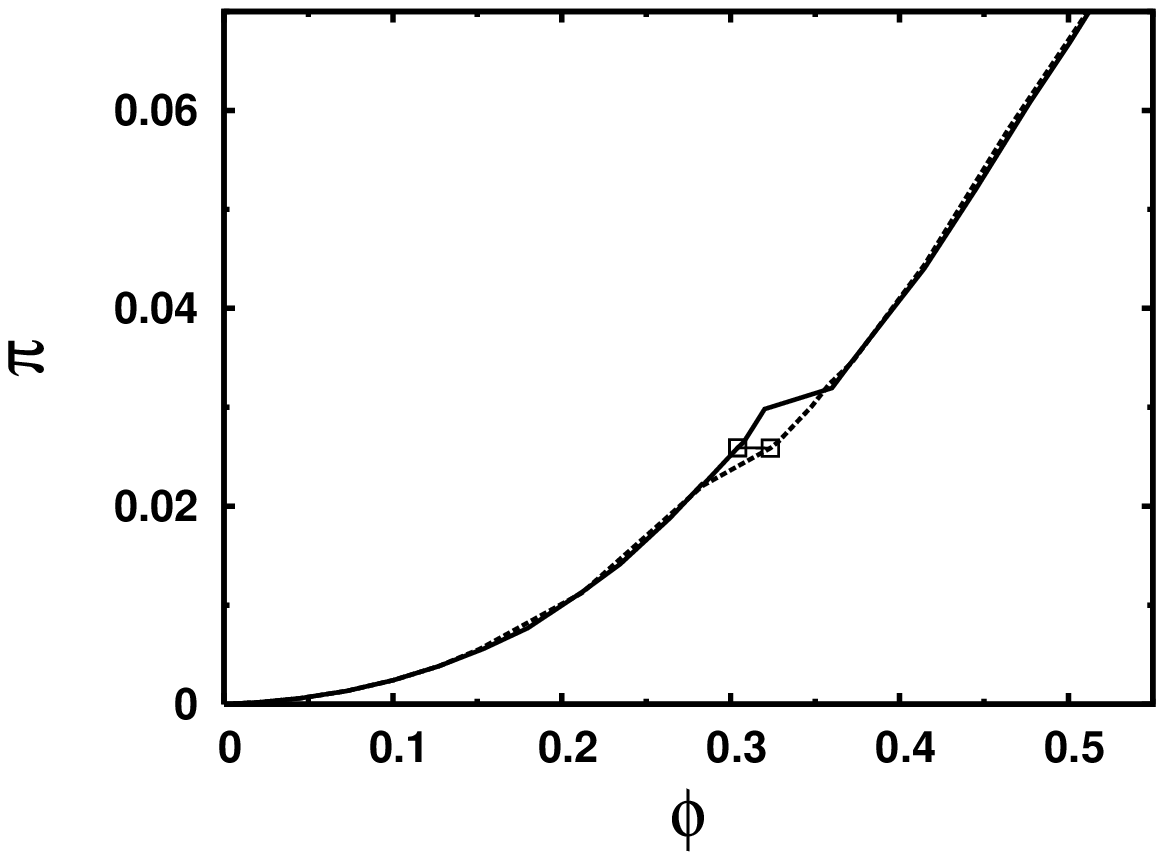}
\caption{(a) The dependence of the pressure vs. chemical potential for
$f=8$, $\varepsilon_0=0$, obtained by TI$\mu VT$ method; the inset
shows the difference $\pi_{nematic}-\pi_{isotropic}$ in the region
close to the isotropic-nematic transition on enlarged scales. Filled
triangles correspond to a dense packed starting conformation,
while open squares to a dilute isotropic one. (b) The equation of state for
$f=8$, $\varepsilon_0=0$ (solid and dotted lines).
The hysteresis region as well as determined
transition line are well visible. Two open squares indicate densities
in the coexisting phases.
\label{Fig16}}
\end{figure}

The method to locate the isotropic--nematic transition is shown in Fig.~\ref{Fig16}.
Again, the isotropic and nematic branches $\pi_{iso}(\mu)$, $\pi_{nem}(\mu)$ in Fig.~\ref{Fig16}a
meet in the transition region (values of $\mu$ between $-170$ and $-160$) almost tangentially.
Nevertheless, the intersection point of these two branches can be determined
with a very good accuracy. The inset in Fig.~\ref{Fig16}a shows the difference
$\pi_{nem}(\mu) - \pi_{iso}(\mu)$ in the hysteresis region. This curve crosses zero at
the value $\mu \approx -166$. The statistical error
in this region was less than 0.5\%, therefore we have indicated the 1\% error bars
for all data points in the inset. Additionally, we have found the intersection point
of linear fits of both branches within and in the close vicinity of the hysteresis region
(their slopes are quite close to each other but still different). From these two procedures
of data analysis we were able to determine the transition point as $\mu_{tr}=-166 \pm 0.5$
and $\pi_{tr}=0.026 \pm 0.001$.
The value of $\mu_{tr}$ is indicated in Fig.~\ref{Fig13},
and that of $\pi_{tr}$ indicates the transition in
Fig.~\ref{Fig16}b where also the densities in coexisting isotropic and nematic phases
(determined from Fig.~\ref{Fig13}a) are shown. Note, that the hysteresis in the equation
of state for this value of chain stiffness is much broader than that in Fig.~\ref{Fig10}.
Apart from the problem, that the finite size of the box in the $x$- and $y$-directions
may still be responsible for some systematic errors (for $L=80$ the size exceeds
$\sqrt{\langle R_e^2 \rangle}$ only by about a factor $2$), an accurate location of
the transition and characterization of the discontinuities is possible.

\begin{figure}[p]
a) \resizebox*{0.65\textwidth}{!}{\includegraphics[angle=0]{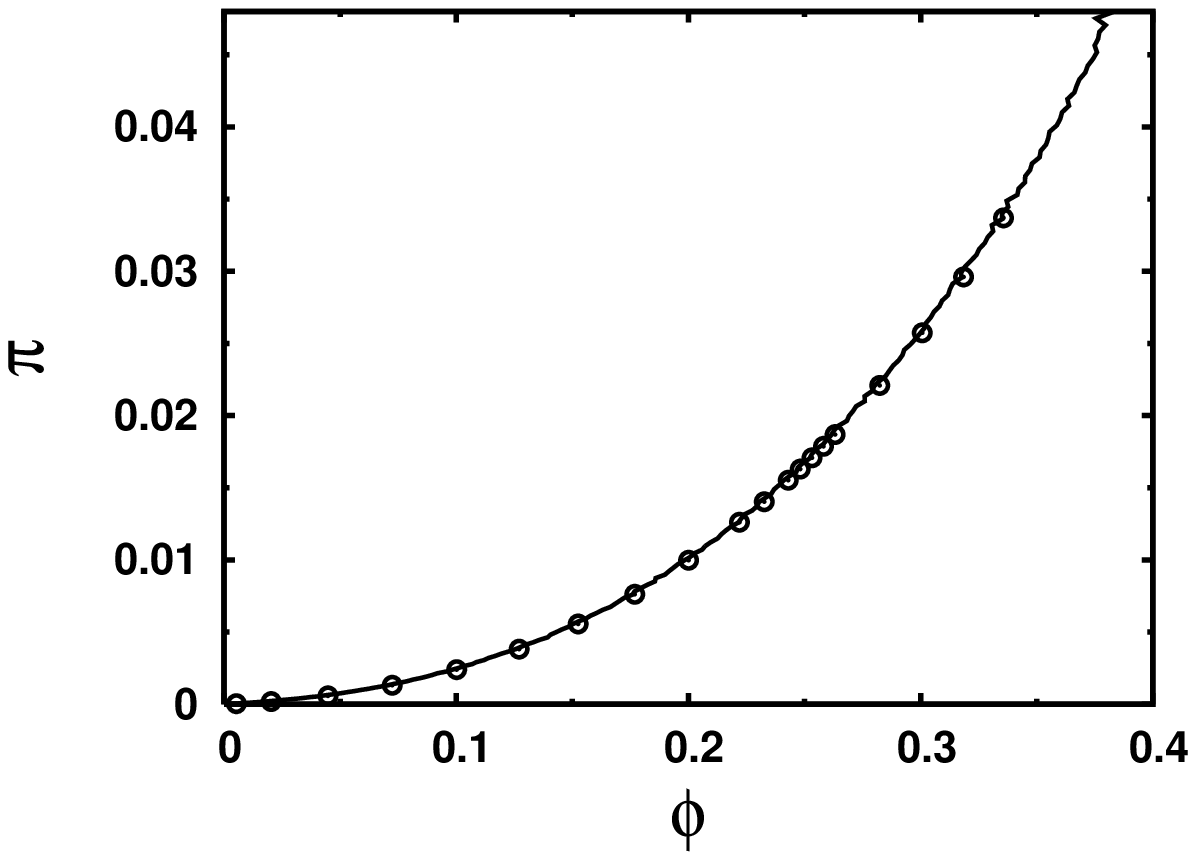}}

b)\resizebox*{0.65\textwidth}{!}{\includegraphics[angle=0]{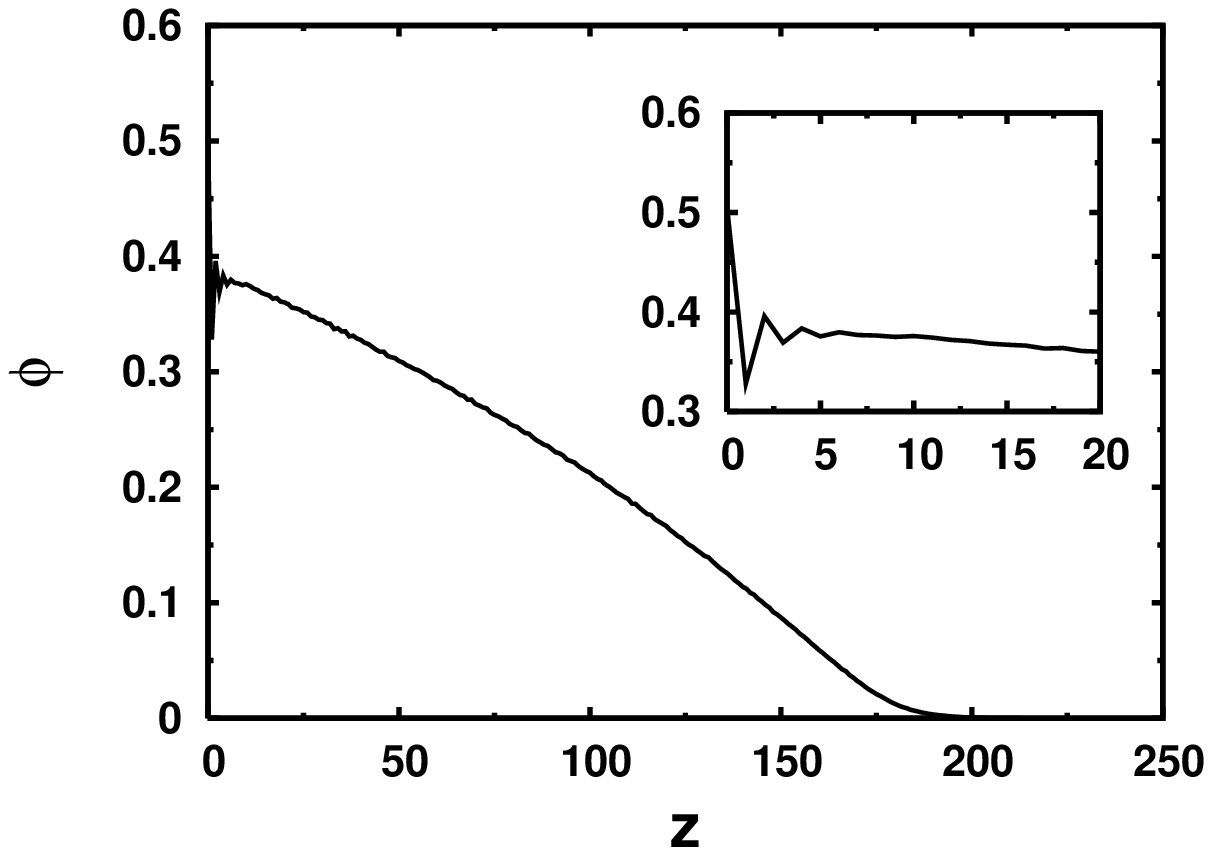}}
\caption{(a) Equation of state $\pi(\phi)$ for
an athermal solution of semiflexible chains, with the parameters
of the bending energy \{Eq.~(\ref{Estiff})\} chosen as
$\varepsilon_0=0$ and $f=4$ (a). Open circles are the results of
the TI$\mu VT$ method, while the solid line is the result of the SE
method, choosing the potential Eq.~(\ref{ExtPot}) and
$\lambda_g=0.01$. The box size was equal to $80\times 80\times 250$,
the number of chains was equal to ${\cal N} = 1600$. Part (b)
shows the density profile for the SE method. The inset shows an
enlarged region in the vicinity of the wall where the layering
is well visible.
\label{Fig12}}
\end{figure}

Now we turn to the test of the SE method for solutions of semiflexible
chains (Fig.~\ref{Fig12}). It is found (see Fig.~\ref{Fig12}a)
that for the case $f=4$, $\varepsilon_0=0$ the SE method is in very
good agreement with the TI$\mu VT$ method, as in the case of flexible chains.
Note however, that the regime of parameters studied here does not yet
encompass the nematic phase.
Similar good results are found for the case $f=2.68$, $\varepsilon_0=4$
discussed in Fig.~\ref{Fig10} (to save space these data are not shown here).

\begin{figure}[p]
a) \includegraphics[width=0.7\textwidth,angle=0]{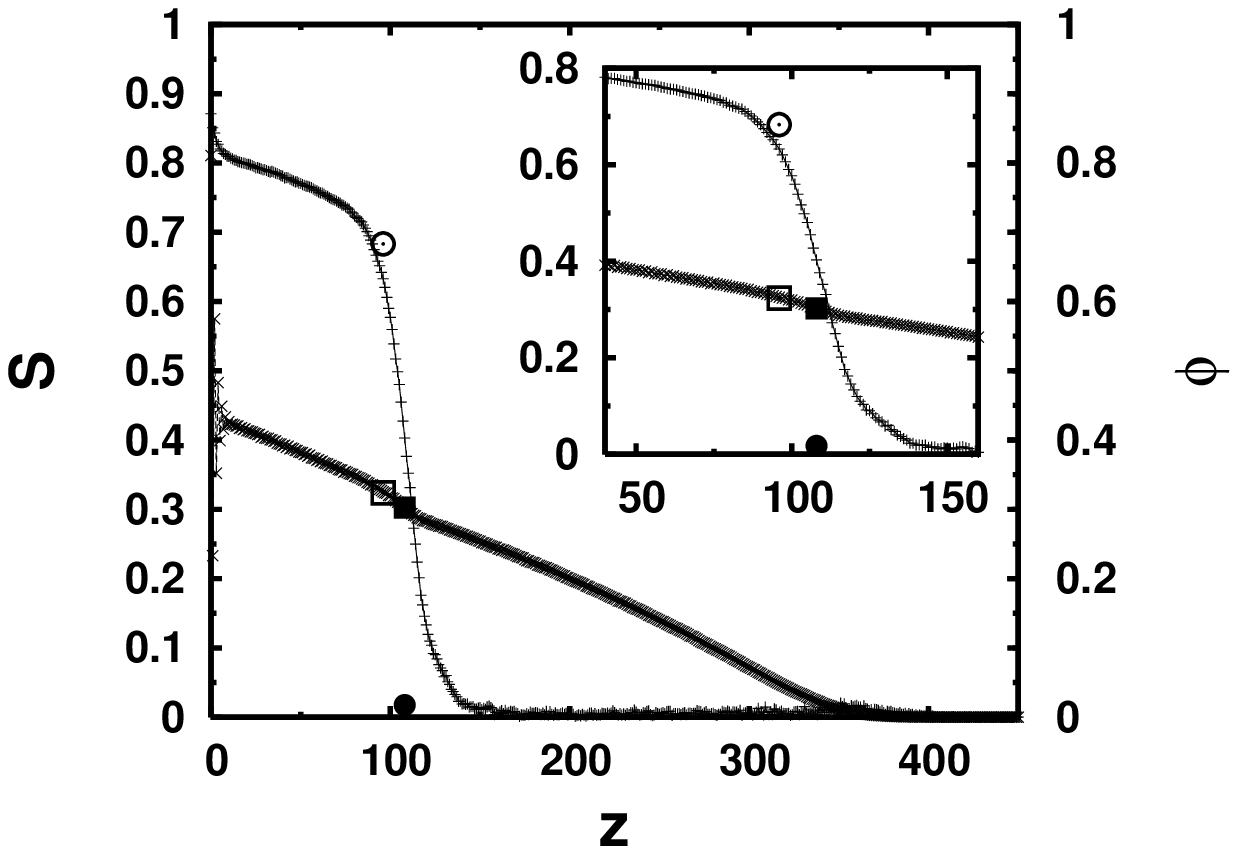}

\vspace{1cm}

b) \includegraphics[height=0.2\textwidth,angle=0]{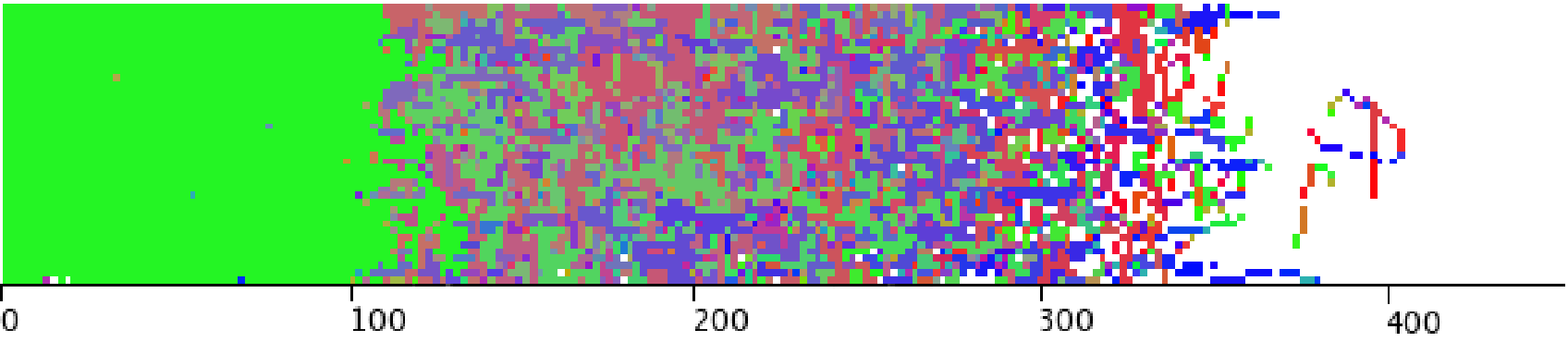}
\caption{(Color online) (a) Profiles of the orientational order parameter,
$S(z)$ (pluses $+$), and of the volume fraction of monomer units,
$\phi(z)$ (crosses $\times$), for the $80\times 80\times 1000$ system, for the
parameters $f=8.0$, $\varepsilon_0=0$, $N=20$, ${\cal N}=1600$,
$\lambda_g=0.01$. Squares indicate the density values at coexistence,
while the circles indicate the values of the nematic order parameter
at coexistence. These values were extracted from the bulk grand
canonical simulation (Fig.~\ref{Fig13}). The inset shows an enlarged
region with the isotropic-nematic interface.
(b) Two dimensional $xz$-map of the
coarse--grained order parameter profile for a system snapshot
corresponding to (a); for details of calculation see the text and
Ref.\onlinecite{LCpaper2}.
\label{Fig17}}
\end{figure}

However, for the case $f=8.0$, $\varepsilon_0=0$ some problems of SE method start to
appear because of the formation of a nematic phase formed on the hard wall.
What happens in the system is shown in Fig.~\ref{Fig17}a
where the profile of the orientational order parameter, $S(z)$, is plotted together with
the density (volume fraction) profile, $\phi(z)$. The standard nematic order parameter tensor,
Eq.(\ref{Qtensor}), was calculated for bond vectors
in each of the $xy$-layers along the $z$-axis separately and averaged over many conformations.
Afterwards its maximal eigenvalue was calculated giving the orientational
order parameter $S$ in each layer.
The points indicated by squares and circles in Fig.~\ref{Fig17}a (and
in the inset showing the transition regime in an enlarged scale) present the data for
density (squares) and orientational order parameter (circles) in the isotropic (filled symbols)
and in the nematic (open symbols) phases obtained from the grand canonical
simulations at $\mu =-166$ (see Fig.~\ref{Fig13})
which exactly corresponds to the transition point.
We used the values of the bulk densities to locate the layer $z$ where
the same value of density occurs in the system with the wall, and then
we plotted the value of $S$ in the bulk system at the same layer
$z$. In Fig.~\ref{Fig17}b we present the two-dimensional $xz$-map of
the coarse-grained order parameter profile for this system using red,
green and blue colors to represent the average local orientation of
monomer units along $x$, $y$ and $z$ axes, respectively (the details
of the calculation method can be found in
Ref.\onlinecite{LCpaper2}). From both these figures (Fig.~\ref{Fig17}a
and~\ref{Fig17}b) one can see that the wall stabilizes rather thick
domains of the nematically ordered phase.

One can observe a kink in the z-dependence of the density profile
exactly between the two coexisting bulk densities of the
isotropic-nematic transition. The order parameter, however, starts to
rise significantly before the limiting isotropic density in the bulk
is reached and has a value around $0.5$ at this density, whereas the
corresponding bulk value is close to zero. The order parameter profile
appears strongly rounded and slightly shifted with respect to the bulk
simulations. These effects prevent an exact localization of the
values of density and order parameter at the isotropic-nematic
transition by the SE method. The rounding of the transition is
unavoidable and mainly due to the presence of capillary wave excitations
of the interface\cite{rowl-widom}, which would even increase in magnitude upon an
increase of the lateral dimension of the simulation box\cite{capill01,capill02},
in contrast to the grand canonical bulk simulations where one can reduce finite
size effects by increasing the system size. The slight shift of the
transition in the order parameter profile as opposed to the density
profile is an indication of a precursor of a nematic wetting layer at the hard
wall.

\begin{figure}[p]
\includegraphics[width=0.65\textwidth,angle=0]{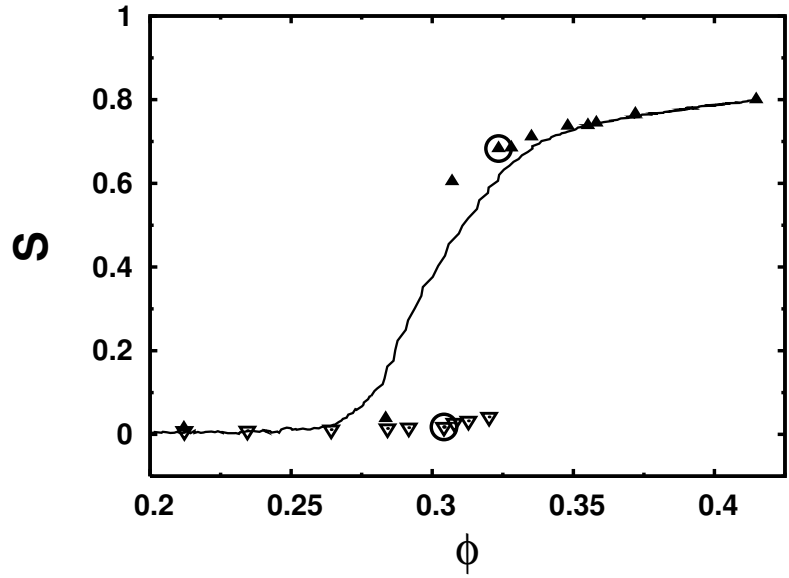}
\caption{Dependence of the nematic orientational order parameter $S$ on the density $\phi$:
comparison of TI$\mu VT$ data (open and filled triangles) with SE data (solid line). The
two large open circles indicate the isotropic--nematic transition.}
\label{Fig18}
\end{figure}

It is interesting to compare the dependence of the orientational order parameter $S$
on the density $\phi$ (see Fig.~\ref{Fig18}) obtained by means of SE and TI$\mu VT$ methods
which can be extracted from the data presented in Fig.~\ref{Fig17}a and Figs.~\ref{Fig13}a
and~\ref{Fig13}c, respectively. Again, the TI$\mu VT$ data show a hysteresis while
the SE data exhibit a smooth transition. It should be emphasized that both in the
isotropic and nematic phases outside the hysteresis region the curves for both methods
coincide with each other
indicating that the SE method reproduces the properties of nematic phase correctly despite
the vicinity of a hard wall.

\begin{figure}[p]
a) \includegraphics[width=0.6\textwidth,angle=0]{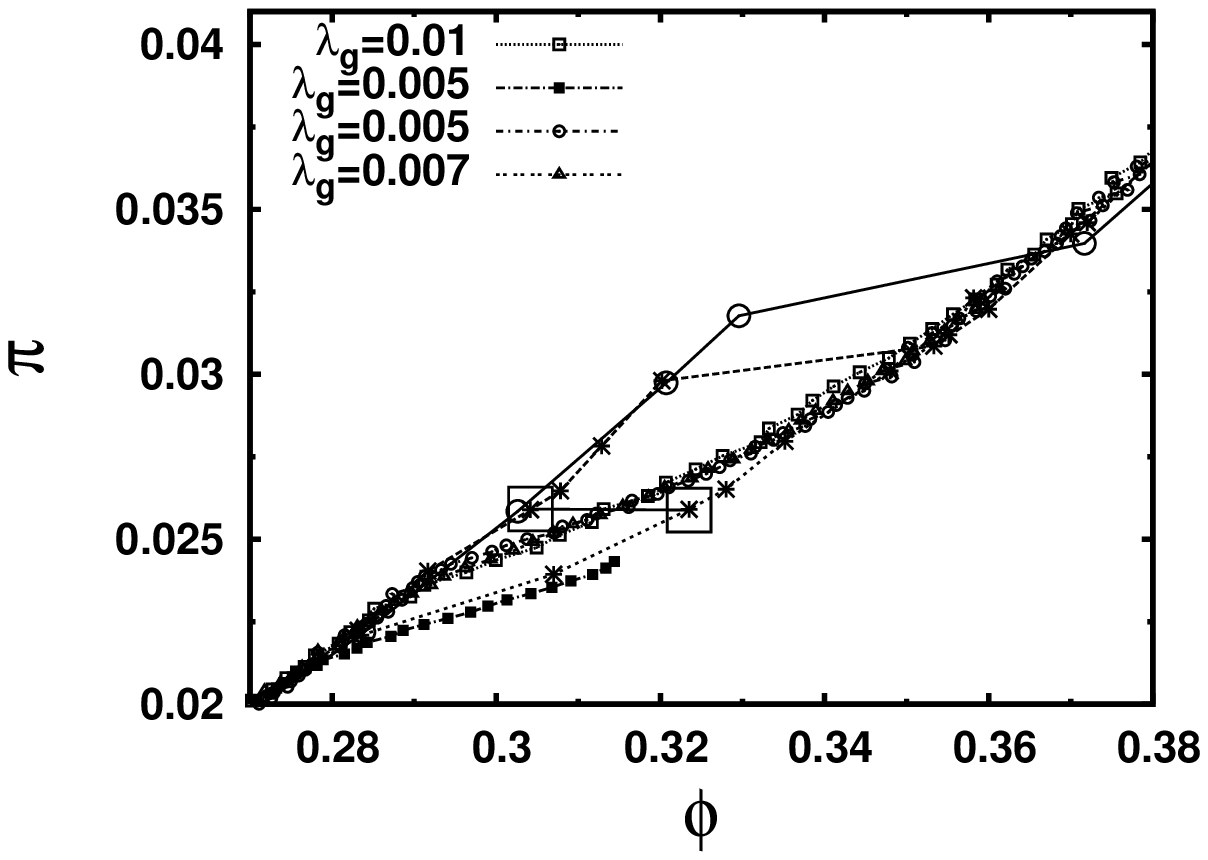}

b) \includegraphics[width=0.55\textwidth,angle=0]{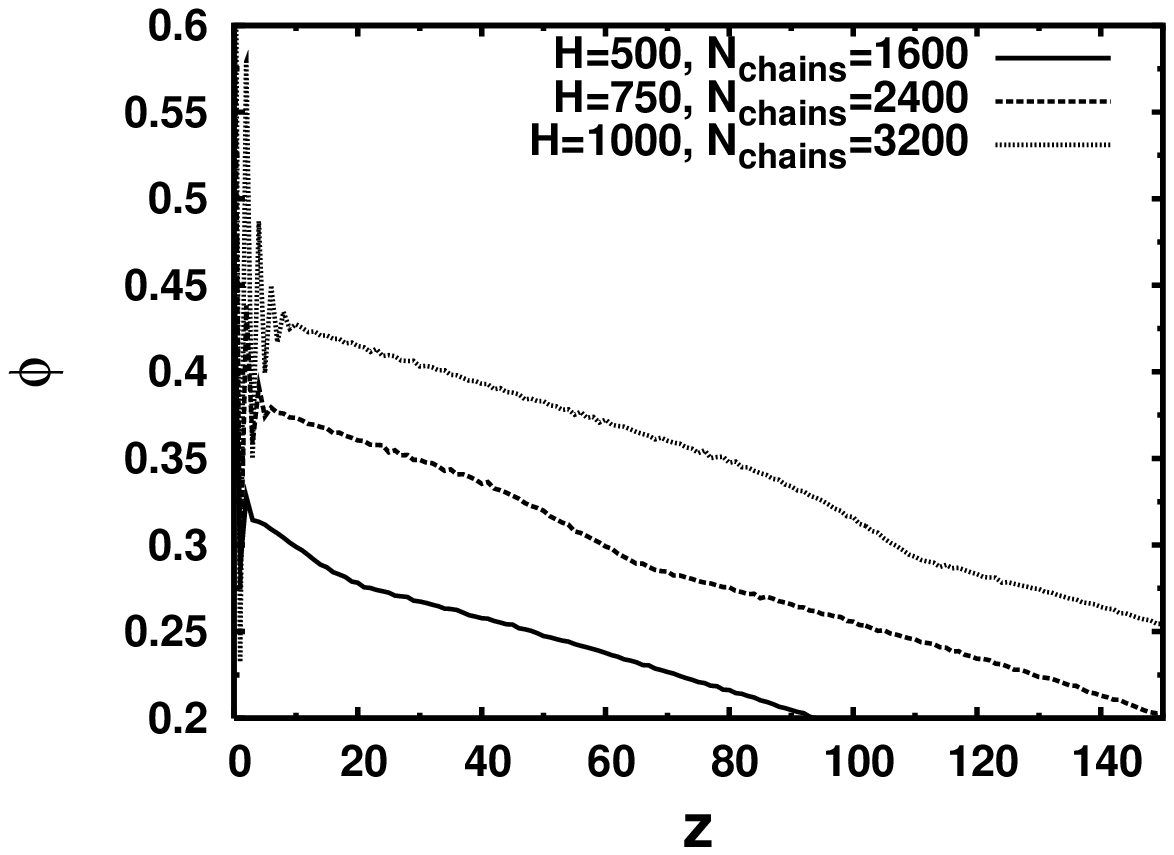}

\caption{(a) Equation of state $\pi(\phi)$ for the case $f=8.0$,
$\varepsilon_0=0$, $N=20$, $L=80$ using the SE method for several choices
of ${\cal N}$, $H$ and $\lambda_g$ as indicated in the legend:
$\lambda_g=0.01$, $H = 250$, ${\cal N}=1600$ (small open squares);
$\lambda_g=0.005$, $H = 500$, ${\cal N}=1600$ (small filled squares);
$\lambda_g=0.005$, $H = 1000$, ${\cal N}=3200$ (small open circles);
$\lambda_g=0.007$, $H = 500$, ${\cal N}=1600$ (small filled triangles).
Additionally, large open circles (and the solid line which is only a guide
for the eye) show the result of the TI$\mu VT$ method
for the box $90 \times 90 \times 90$.
Stars and dashed lines (as a guide for the eye) are the
results of the TI$\mu VT$ simulations in the box $80 \times 80 \times 150$
(from Fig.~\ref{Fig16}b).
The isotropic--nematic transition line is visible between
two large open squares indicating the coexisting densities.
Regions of the density profiles with a kink indicating the
isotropic--nematic transition are presented in (b) for
$\lambda_g=0.005$ and different $H$ and ${\cal N}$ as indicated in
the legend.
\label{Fig14}}

\end{figure}

The inevitable presence of the interface leads to a
smoothing of the first order transition on the pressure vs. density dependence
(Fig.~\ref{Fig14}a). In this Figure both the data for the SE method are presented
using different values of $H$ and $\lambda_g$, as well as the data for
the TI$\mu VT$ method which we have already discussed above.
It is clear from the principle of the SE method that it is impossible to
observe any hysteresis on the equation of state.
Fig.~\ref{Fig14}a shows that the $\pi(\phi)$ curves generated by the SE method
start to deviate from those generated by TI$\mu VT$ method
near $\phi=0.28$, while for $\phi<0.28$ the
agreement between both methods is excellent as well as for $\phi>0.36$
(these parts are not shown here).
All curves for the SE method obtained at different values of $H$ and $\lambda_g$
almost coincide with each other (a variation of the lateral box linear
dimension $L$ has no detectable effect on SE results)
except for one data set shown with small filled squares
($H=500$, ${\cal N}=1600$, $\lambda_g=0.005$).
In Figs.~\ref{Fig14}b the density profiles are presented and the
reason for the deviation of the one data set becomes apparent: this is the only
curve where the isotropic--nematic interface (which shows up as a characteristic kink
in the density profile) appears very close to the hard wall (note that
for $f=8.0$ the end-to-end distance of a chain is $R_e \approx 40$
(Fig. 8) and the chains are already rather stretched (Fig. 12b))
so that layering effects influence the isotropic--nematic coexistence
in this case significantly. The effect of the layering at the wall, which can be quite strong
(see Fig.~\ref{Fig14}b) on the equation of state in the isotropic
regime decreases with increasing distance of the isotropic-nematic
interface from the wall, such that the curves for the SE method in
(Fig.~\ref{Fig14}a) already coincide within our error bars.
Note also, that the ordinate of the kink in the density profiles in
Fig.~\ref{Fig14}b is the same for all systems
at different parameters. For the bulk simulation (the TI$\mu VT$ data) for different
systems presented in Fig.~\ref{Fig14}a we can conclude that
there exists an incommensurability effect which influences the equation of state: large
open circles show the well equilibrated data obtained in the cubic box $L=90$, $H=90$
(box 1, see also above), and the hysteresis region is different from the one
obtained for $L=80$, $H=150$ (box 2) shown with stars and large open squares.

\begin{figure}[p]

\includegraphics[width=0.65\textwidth,angle=0]{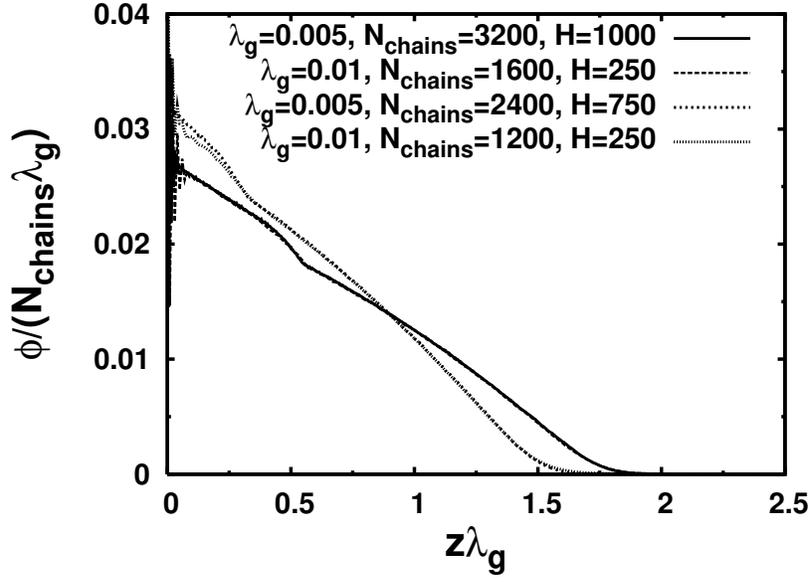}
\caption{ Rescaled density profiles $\phi/({\cal N}\lambda_g)$
vs. $z\lambda_g$ for several systems, as
indicated in the legend. Note that curves
with the same value of ${\cal N}\lambda_g$ practically
superimpose. The average value of the gyration radius
$\sqrt{\langle R_g^2 \rangle}$ is about $14.8$ both in the dilute
and concentrated regions.
\label{Fig15}}
\end{figure}

Finally, we mention here a scaling of the density profiles shown in Fig.~\ref{Fig15}
(for details see the figure caption) which is also in agreement with results
obtained in Ref.\onlinecite{Hansen}. The curves superimpose for the systems where the
combination of parameters ${\cal N} \lambda_g$ has the same value. The curves for
larger values of ${\cal N} \lambda_g$ look broader and smoother in these scaled variables
$\phi / ({\cal N} \lambda_g)$ vs. $z \lambda_g$ and show up the kink (isotropic--nematic
interface) further away from the hard wall.

It is interesting to compare our conclusion on the
shift of isotropic--nematic transition with the results of computer simulation
studies of hard rod colloidal suspensions in confinements, i.e. solutions confined
between hard walls and/or exposed to an external gravitational
potential. A shift of the isotropic--nematic transition to lower densities as compared to
the bulk was found for a hard-rod fluid confined by two walls\cite{Dijkstra03}. At the same time,
a surprisingly good agreement between two osmotic equations of state for hard-rod
fluids obtained from computer
simulation using the SE method and from bulk simulations at many different densities has
been reported in Ref.\onlinecite{Dijkstra04}, also for densities in the nematic phase.
The authors of Ref.\onlinecite{Dijkstra04} explained this agreement by a very small
interfacial width of the isotropic--nematic interface in comparison with the
gravitational lengths considered in their work, a situation which is
also realized in our simulations. However, for the model in
Ref.\onlinecite{Dijkstra04} the density difference between isotropic
and nematic phase is very small and any possible deviations between the
bulk equation of state and the results from the SE method simulations
in the crossover density regime are not visible within the resolution
and statistical uncertainty of the simulations.


Theoretically the effect of gravity on the phase behavior of hard rod solutions was
studied in Ref.\onlinecite{Baulin}. Comparing density and order parameter profiles
in Figs.~\ref{Fig17}a with those calculated in Ref.\onlinecite{Baulin} we can see
a very good agreement: the density profiles show a quite small jump at the transition
point and a smooth but significant decrease both in nematic and isotropic phases,
while the orientational order parameter is almost constant within each of two phases
and experiences a quite large jump at the transition point.

\section{Conclusions}
\label{conclusions}

Monte Carlo computer simulations using the bond fluctuation model have been performed
for solutions of semiflexible chains of the length $N=20$ monomer units.
Three methods for pressure calculation in lattice Monte Carlo simulations have
been investigated and compared:
(1) the thermodynamic integration method in the grand canonical ensemble (TI$\mu VT$);
(2) the repulsive wall method in the grand canonical ensemble (RWTI);
(3) the sedimentation equilibrium method (SE)
in the canonical ensemble in an external sedimentation field.
All three methods show quite similar results
for solutions of flexible chains as well as for the region of the isotropic phase
of semiflexible chains. However, differences may occur at higher densities (or pressures)
where for semiflexible chains the transition to the nematic phase takes place.
Methodological problems of pressure measurement in solutions in the vicinity of
isotropic-nematic transitions have been discussed.

The most crucial point is that the presence of a hard repulsive wall and/or an
external sedimentation field exerts a significant influence on the isotropic-nematic
transition, both on the transition point (transition density) and also on the structure
of the ordered phase.

Thus we have found that the SE method is useful for obtaining the equation of state
of various polymeric systems but its use becomes problematic
in the vicinity of phase transitions. We have demonstrated this for the isotropic-nematic
transition in solutions of semiflexible chains. The SE method works
quite well sufficiently below  and above
the density of the isotropic-nematic transition, but it fails to predict the transition density
correctly, at least for system sizes (which were reasonably large) and sedimentation field
strengths (which were reasonably small) used in our simulations.

The source of the problems that we encountered is that for a system undergoing a transition
from the isotropic to the nematic phase the SE method implies that for densities
large enough that the nematic phase can develop in the simulation box, one necessarily must have
a transition zone of densities where the nematic--isotropic interface is present in the box.
This nematic--isotropic interface is not sharp but rather extended, and hence it is not clear
from data such as Fig.~\ref{Fig17} to judge where the region of the ''bulk'' nematic phase
stops and where the region of the interface begins. In fact, for an equilibrium interface in the
absence of an external (gravitational) field we would have a smooth interfacial profile
between the coexisting phases as well. This profile can be (at least approximately) considered
as the convolution of an intrinsic profile with capillary--wave--induced broadening, which
increases with the logarithm of the lateral system size, proportional to $\ln L$.
Note, that already in this case the problem of disentangling the interfacial profile
from the capillary wave broadening is notoriously difficult. However, in the presence
of a gravitational field the problem is even more subtle: while strong gravitational fields
will lead to a ''squeezing'' of the intrinsic interface (similar to the squeezing
of interfaces by very strong confinement), and eliminate the capillary wave broadening
completely, weak gravitational fields will leave the intrinsic profile more or less intact,
but eliminate the long wavelength part of the capillary wave spectrum. As a consequence,
one expects a crossover from a broadening proportional to $\ln L$ for not too large $L$ to
a finite width independent of $L$ but controlled by the strength of the gravitational field.
An explicit study of all these interfacial phenomena is beyond the scope of the present paper.
The problem gets even more complicated by the fact that also in the ''bulk'' nematic
phase the nematic order parameter is not at all constant, but varies with the distance $z$ from
the wall, since the order parameter near the transition depends strongly on density, and thus
a strong variation of order parameter is caused by the density variation as well.
In view of these problems, it is difficult from the SE method to estimate accurately at which density the region
of the isotropic phase stops and at which (higher) density then the region of the nematic
phase begins. An estimation of the nematic order parameter at the transition point is hardly
possible.

A particularly subtle difficulty occurs if parameters such as system size $H$ in the long
direction and strength of the gravitational potential are chosen such, that there is not enough
space left for the nematic phase to develop, and the isotropic--nematic interface occurs
directly adjacent to the wall (Fig.~\ref{Fig14}b). Then data for the osmotic pressure in the
transition region are obtained that are systematically too small, and suggest a transition
from the isotropic to the nematic phase at a density that is clearly too low. This example
shows that one must not rely on the SE method blindly when a phase transition occurs, but one
needs then to check that the results are not changing when the strength of the gravitational
field and/or the size in the long direction are varied.

We restricted ourselves to the chain length $N=20$ monomer units because in the
bond-fluctuation model this chain length is sufficiently long to display Gaussian
behavior in its chain statistics in the melt (i.e., it is a real flexible polymer,
not an oligomer) and it is not too long to allow for a sufficiently efficient simulation.
We should emphasize that the SE method is valid for polymer chains of any
length, provided the length scale of variation of the external potential in which
the sedimentation equilibrium is reached is much larger than the radius of gyration
of a polymer chain, e.g., flexible chains of the length up to 500 monomer units
have been studied in Ref.\onlinecite{Hansen}. However, for the semiflexible chains the
effect of chain stiffening in the nematic state\cite{LCpaper1} will necessarily
require much larger simulation boxes in comparison to the case of flexible chains of the
same length.

While the TI$\mu VT$ method near the isotropic--nematic transition is hampered by hysteresis,
Fig.~\ref{Fig13}, the fact that the transition in the $(\pi, \mu)$ plane must show up as
a simple intersection point of the curves does allow an accurate location of the transition
(Fig.~\ref{Fig16}a), and to characterize the magnitudes of the jumps. In this way, the
strictly horizontal part in the $\pi$ vs. $\phi$ isotherm (Fig.~\ref{Fig16}b) can be constructed.
Thus, we feel that for the accurate characterization of first order phase transitions the
TI$\mu VT$ method is preferable, whenever applicable. With respect to the numerical data
presented in this paper, we add the caveat that there may be still some systematic errors due
to a too small value chosen for the lateral size $L$. However, the same size was used for
the simulations by means of SE method, and hence we think that our discussion of the relative
merits of these methods should not be affected by this problem.

In future work, we plan to extend these studies to include also attractive interaction
between the monomer units, to investigate the competition between nematic ordering and
polymer--solvent phase separation, applying the methods validated in the present paper.

\section*{Acknowledgments}
We acknowledge useful discussions with Dr. T. Schilling and
the financial support from DFG (grant 436 RUS 113/791),
RFBR (grant 06-03-33146), and Alexander-von-Humboldt-Foundation.

\newpage

\end{document}